\documentclass[a4paper,notitlepage,11pt]{article}
\newcommand{\rh}[1]{$\rho = 10^{-#1}\sigma^{-3}$}
\newcommand{\ro}[1]{$\rho = 2 \cdot 10^{-#1}\sigma^{-3}$}

\newcommand{\lb}{$\lambda_{B}$\enspace}
\usepackage[dvips]{epsfig}
\usepackage{amstex}
\usepackage{amssymb}
\begin{document}
\begin{titlepage}
\begin{center}
\Large {\bf
Strongly Charged, Flexible Polyelectrolytes \\[0.2cm]
in Poor Solvents -- \\[0.2cm]
Molecular Dynamics Simulations \\[0.5cm]}
\large {\em Uwe Micka, Christian Holm, and Kurt Kremer\footnotemark\\
\large Max-Planck-Institut f{\"u}r Polymerforschung \\
\large Ackermannweg 10 \\ 55128 Mainz, Germany \\[0.4cm]}
August 28, 1998\\
\end{center}
\footnotetext{Author to whom correspondence should be addressed.}

\begin{abstract}
We present a set of molecular dynamics (MD) simulations of strongly charged,
flexible 
polyelectrolyte chains under poor solvent conditions in a salt free solution. 
Structural properties of the chains
and of the solutions are reported. 
By varying the polymer density and the electrostatic 
interaction strength we study the crossover
from a dominating electrostatic interaction to the regime of strong screening, 
where the hydrophobic interactions dominate. During the crossover a multitude
of structures is observed. 
In the limit of low polymer density strongly stretched, necklace like 
conformations are found. In the opposite limit of high polymer density which 
is equivalent to strongly screened electrostatic interactions, we find that 
the chains are extremely collapsed, however we observe no agglomeration or 
phase separation.
The investigations show that the density of free charges is one of
the relevant parameters which rules the behavior of the system and 
hence should be used as a parameter to explain experimental results.
\end{abstract}
\verb+E-mail: kremer@mpip-mainz.mpg.de, holm@mpip-mainz.mpg.de+
\end{titlepage}

%
\section{Introduction}
%
Unlike neutral polymers \cite{flory,deGennes,DoiEdwards},
the understanding of the behavior of electrically charged macromolecules,
short polyelectrolytes, is still rather poor. The long range nature of the
electrostatic interactions
introduces new length and time scales that render the analytical description
very complicated \cite{JoBaReview} and prevent simple scaling arguments
to work equally successfully as in the neutral case. Usually, analytic theory
\cite{tanni} assumes $\theta$-properties for the uncharged monomers, while
simulations use either $\theta$- or good solvent conditions 
\cite{paper1,Mark1,Markrest,paper2,JoBa}.
Nevertheless, most experimentally relevant systems have organic, non-polar
backbones in polar solvents (water) so that poor solvent conditions are more
appropriate \cite{FoersterSchmidt}. The analytic description of these 
systems is not far developed. In a first attempt, Dobrynin et al. \cite{neckl} 
studied a single weakly charged chain under poor solvent conditions. 
Starting from the description of a charged
drop, originally given by Lord Rayleigh in 1882 \cite{Ray},  their theory
predicts that when the charge density or the strength of
the electrostatic repulsion increases, the globular state becomes unstable and
starts to split into smaller globules, which try to separate.
Due to the covalent bonds the parts cannot dissociate completely. By 
maximizing their distance, they form structures which are called
pearl-necklace chains.

The approach of Dobrynin et al. is supported by a Monte Carlo simulation of a
single chain that shows a cascade from one to two and three globules with
increasing strength of the electrostatic repulsion \cite{neckl}. In their
study every
monomer carries a fractional charge, the polyelectrolyte is weakly charged and
no explicit counterions are taken into account. 
Additionally to this previous Monte Carlo study, there exist 
some experimental evidence \cite{willi2} supporting the picture of Dobrynin at
al.   
A recent Monte-Carlo simulation
of single Debye-H\"uckel chains near the $\theta$ transition also supports 
these findings \cite{duenweg98}. However, there has been also put some doubt
on the validity of this scheme in the case of discrete ions \cite{morawetz}.
Our study will use many chains with discrete counterions, where the charges
all interact via the full Coulomb potential, simulated wit molecular
dynamics (MD).

Alternatively, a cylindrical instability was proposed by Khokhlov and his
collaborators \cite{KoKa}. They
argue that there will be a micro-phase separation due to a kind of
agglomeration of the collapsed chains. This state would increase the entropy
of the condensed counterions which could explore the whole volume of the
phase separated chains.  
In an earlier short communication
\cite{letter} we showed that based on simulational data 
no phase separation was observed and gave a simple argument 
for a stable colloid like state of single collapsed chains.
In this article we will present in detail our simulations of
polyelectrolytes in poor solvents. The rest of the paper is organized as
follows: 
In
section \ref{model} we will describe our model, and show in section
\ref{ww} the strong influence of the hydrophobic interaction, when it is put
into a right proportion to the electrostatic interaction.
In section \ref{konf} we will present the single chain properties of a
polyelectrolyte under poor solvent condition, which shows indeed a cascade of
pearl-necklace structures, if one varies the polymer density. We also explain
why the colloid like state of the collapsed chain seems to be the stable one,
and why no phase separation is observed.
As last point in section \ref{exp} we compare our results to experimental data
and suggest for future experiments those parameters 
which we consider to be the important quantities to vary.
%
\section{Model \label{model}}
%

The polyelectrolyte chains are modeled as bead spring chains of Lennard-Jones
(LJ) particles.

For good solvent conditions, a shifted LJ potential is used to describe the
purely repulsive excluded volume interaction between all monomers:
\begin{equation}\label{labelgleich}
U_{LJ,hydrophil}(r) = 
\begin{cases}
4\cdot \epsilon_{LJ} [(\frac{\sigma}{r})^{12} - (\frac{\sigma}{r})^{6}] +
\epsilon_{LJ} & \text{for } r \le r_{min}=2^{\frac{1}{6}}\sigma \\
\hspace*{0.95cm} 0 & \text{for }  r > r_{min}=2^{\frac{1}{6}}\sigma
\end{cases}
\end{equation}
The parameter $\sigma$ is used to set the length scale.
The energy parameter $\epsilon_{LJ}$ controls the strength of the interaction,
and its value for the good solvent case is fixed to 
$\epsilon_{LJ} = 1.0 k_B T$, where $k_B$ is the Boltzmann constant and $T$ is
the temperature.
Polymers that are in good solvent and whose monomers 
interact via this potential are called ``hydrophilic'' for the rest of this
paper. Although we do not have any explicit solvent and no
specific potentials for the solvent water, we use this
abbreviation because water is in experiments, technical applications, and
nature the most 
common solvent for polyelectrolytes. As a consequence, particles under poor
solvent conditions are called ``hydrophobic''. 
Because we do not consider any explicit interactions between solutes and
solvent molecules one way to model poor solvent conditions is through an
additional attractive interaction between the chain monomers.

We represent this potential by a standard LJ potential with a larger cut-off
($R_{c}=2.5 \sigma$):
\begin{equation}\label{ljpotphob}
U_{LJ,hydrophob}(r) = 
\begin{cases}
4\cdot \epsilon_{LJ} [(\frac{\sigma}{r})^{12} - (\frac{\sigma}{r})^{6}
- c(R_c)]  
& \text{for } r \le R_{c}  \\[0.1cm]
\hspace*{0.95cm} 0 & \text{for }  r > R_{c} \\
\end{cases}
\end{equation}
The function $c(R_c)$ is chosen as 
$c(R_c) = (\frac{\sigma}{R_c})^{12} - (\frac{\sigma}{R_c})^{6}$ to give a
potential value of zero at the cutoff.
This potential is used only for chain beads, because 
we assume that the counterions do not have any hydrophobic parts in their
interaction, as is reasonable for alkali metals, and describe their
interaction by equation
\ref{labelgleich}.

The connectivity of the bonded monomers 
is assured by the FENE (\underline{f}inite
\underline{e}xtension \underline{n}onlinear \underline{e}lastic) potential:
\begin{equation}
U_{\rm FENE}(r) = -\frac{1}{2}k_{\rm FENE}R_{0}^{2}
\ln(1-\frac{r^{2}}{R_{0}^{2}}) 
\end{equation}
where $r$ is the distance of the interacting particles.
The parameters are set the following way: the spring constant $k_{\rm FENE} = 7.0
\frac{k_{B}T}{\sigma^{2}}$ and $R_{0} = 2 \sigma$. $R_{0}$ gives the maximum
extension of the bond, at which the interaction energy becomes infinite.  
This potential gives rather stiff bond lengths with fluctuations being smaller
than 10 \%. The nonlinearity allows a very efficient mixing of the different
vibrational modes.

The Bjerrum length is
a measure of the strength of the electrostatic interaction, which is defined
as the
length at which the electrostatic energy equals the thermal energy:
\begin{equation}
\lambda_{B}=\frac{e^{2}}{4\pi\epsilon_{S}\epsilon_{0}k_{B}T},
\end{equation} 
where 
$e$ is the unit charge of the interacting particles, and 
$\epsilon_{0}$ and $\epsilon_S $ are the permittivity of the vacuum and
of the solvent, respectively ($\lambda_B = 7.1${\AA} in water). 
Here we restrict ourselves 
to monovalent species in salt free
solution. The solvent is taken 
into account via the dielectric background, whose properties
influence the form (hydrophil or hydrophob) 
of the interaction potential of the monomers.
Unless mentioned the Bjerrum length is always fixed to
$\lambda_{B}=3.0 \sigma$.

No assumptions concerning screening effects are made,
instead we are using the pure Coulomb energy to calculate the interactions 
between all charged particles:
\begin{equation}
E^{Coul}(r_{ij}) = k_{B}T \frac{\lambda_B q_i q_j}{r_{ij}},
\end{equation}
where $q_i = +1$ for the charged chain monomers,
and $q_i = -1$ for the counterions, which are treated explicitly as charged LJ
particles. 

The central simulation box contained 16 or 32 chains subject to 
periodic boundary conditions. Each chain consisted of 94 monomers.
The fraction $f$ of charged monomers on the chain is $f = \frac{1}{3}$,
meaning that every third monomer carries a unit charge. We also put a charge
on the chain ends, hence the number of charges per chain is always 32.
A velocity Verlet algorithm is used to integrate the equation of
motion.
The system is coupled to a thermostat via a 
standard Langevin equation with a friction term and
a stochastic force. The friction coefficient $\Gamma$
is set to
$\Gamma = \tau^{-1}$ and the time step $\Delta t$ varies between 
$\Delta t = 0.003 \tau$
and $0.0125 \tau$ depending on $\epsilon_{LJ}$, where $\tau$ is the standard
LJ time unit \cite{Mark1}. 
Because of the periodic boundary conditions the Coulomb interaction is
calculated with a fast variant of the Ewald summation \cite{ewald},
the  
Particle-Mesh-Ewald (PME) algorithm \cite{pmeorig} in the
version presented by Petersen \cite{pmegut}. The basic idea is to calculate
the Fourier part of an Ewald summation with a Fast Fourier Transform on a mesh
to speed up the calculation. Our version is more efficient than the highly
optimized version of the Ewald sum of Ref.\cite{Mark1} if more than 800
charges are simulated. The parameters of the PME method were always chosen
such that the resulting errors in the electrostatic forces were well below the
random forces produced by the Langevin heat bath. At the time of the
production of our data this method was the only mesh based Ewald sum, for
which analytic error estimates \cite{pmegut} were available. Note, however,
that meanwhile a better method, also with analytic error estimates, is
available \cite{dh98}.

Our polyelectrolyte systems are strongly charged. By this we mean that the
Manning parameter which is defined as the Bjerrum length divided by
the length per unit charge, $\xi_M = \frac{\lambda_{B}}{3b}$, is close to the
critical value. The critical
Manning parameter, at which counterion condensation is commonly expected
to set in, is equal to $\xi_M^* = 1$ for
monovalent counterions. In our simulation, the average bond length is given
by $b \approx 1.08 \sigma$, leading to $\xi_M \approx 0.927$, so that we are
just below $\xi_M^*$, therefore the expression ``strongly'' charged.

Our model polyelectrolyte system can be mapped to 
the ``standard polyelectrolyte'' sulfonated polystyrene (PSS), as can
be seen for instance from the Manning parameter. A PSS-monomer extends over
roughly $2.5${\AA} and the Bjerrum length in water is about $7.1${\AA} leading to
about one charge per three monomers.
Each chain contains 94 monomers meaning 32 charges and models a molecular
weight of roughly 20.000 $\frac{\text{g}}{\text{mol}}$, if a sodium counterion
is assumed. This molecular weight, though not very large, 
is well within the experimentally analyzed regime \cite{FoersterSchmidt}.

A crucial parameter in the simulations will be the concentration of particles
in the simulation box of length $L$.
No difference will be made between counterions, charged and uncharged monomers
because they are all Lennard-Jones-particles. This concentration will be called
density for the rest of this paper and therefore be denoted by $\rho$. By
varying the box length $L$
we can vary the density from $2\cdot 10^{-1}\sigma^{-3}$  to $2\cdot
10^{-6}\sigma^{-3}$, which is covering more than the whole experimentally
accessible range from very dense to extremely dilute solutions.
%
\section{Influence of the hydrophobic interaction strength \label{ww}}
%
To characterize the hydrophobicity of a polyelectrolyte chain
one has to develop a measure for the
relative strengths of the hydrophobic and the electrostatic interaction. The
strength of the electrostatic interaction is determined by $\lambda_{B}$. The
attractive interaction in Lennard-Jones systems (as for instance neutral
polymers) is usually tuned by variation of the temperature $T$. 
The aim of the present work is to study the influence of the poor solvent with
fixed boundary conditions (constant $\lambda_B$ and constant $T$). 
For the case of polyelectrolytes a variation of the temperature  in the
simulations changes simultaneously the strength of the electrostatics for
fixed $\lambda_{B}$. For this reason,
the hydrophobicity is varied by changing the parameter 
$\epsilon_{LJ}$ in the LJ potential (\ref{ljpotphob}) as an
independent variable. We fix the energy scale
by setting the thermal energy equal to $k_{B}T=1.0$. 
In a simulation this is a common way to separate and isolate effects. We are
well aware of the fact that in experiments a change of temperature
influences in a rather complex manner all relevant quantities.

In order to make contact with neutral polymer theory
the $\theta$ point for the neutral polymer system
has to be determined first. The $\theta$ point is usually defined 
by the temperature at which the second virial coefficient between the chains
vanishes. There
the attractive and repulsive interactions cancel up to second order, so that
the polymers show (almost) ideal random walk behavior. 
For the underlying neutral LJ system only a crude estimate for
the $\theta$ point has been found so far in Ref.~\cite{muratgrest}. 
The temperature
$T_{\theta}^{(n)}$ and the LJ energy parameter $\epsilon_{LJ}^{(n)}$ of
this point were determined as
$T_{\theta}^{(n)} \gtrsim 3.0 \pm 0.2$ for $\epsilon_{LJ}^{(n)} = 1.0$. 
Converted to our notation this is equivalent to
a value of  $\epsilon_{LJ,\theta}^{(n)} \leq \frac{1}{3}$ at a thermal energy
$k_B T = 1$.  
To get a more precise estimate on the location of the $\theta$ point we
therefore employed a scaling approach that was shown to be very successful
for polymers and star-polymers on a face-centered cubic lattice (fcc)
\cite{batoulis}. This approach utilizes that the end-to-end
distance can be expressed as:
\begin{equation}\label{bateq}
R_{end}^{2} = \text{N} \cdot g(\tilde T \cdot \text{N}^{\frac{1}{2}}),
\end{equation}
where $g$ is an unknown scaling function, and $N$ is the number of monomers of
a chain. The 
argument $\tilde T = \frac{T - T_\theta}{T_\theta}$ 
describes the reduced distance to the
critical point. This relation is easily translated to an
$\epsilon_{LJ}$-dependent 
formula. The main point is that for fixed values of the ratio
$\frac{R^{2}_{end}}{N}$ the argument of the scaling function must be also
fixed. 
It follows that
\begin{equation}\label{neutscalequat}
\tilde T \sim \text{N}^{-\frac{1}{2}}
\end{equation}

\begin{figure}[htb]
  \begin{center}
    \leavevmode
  \epsfig{file=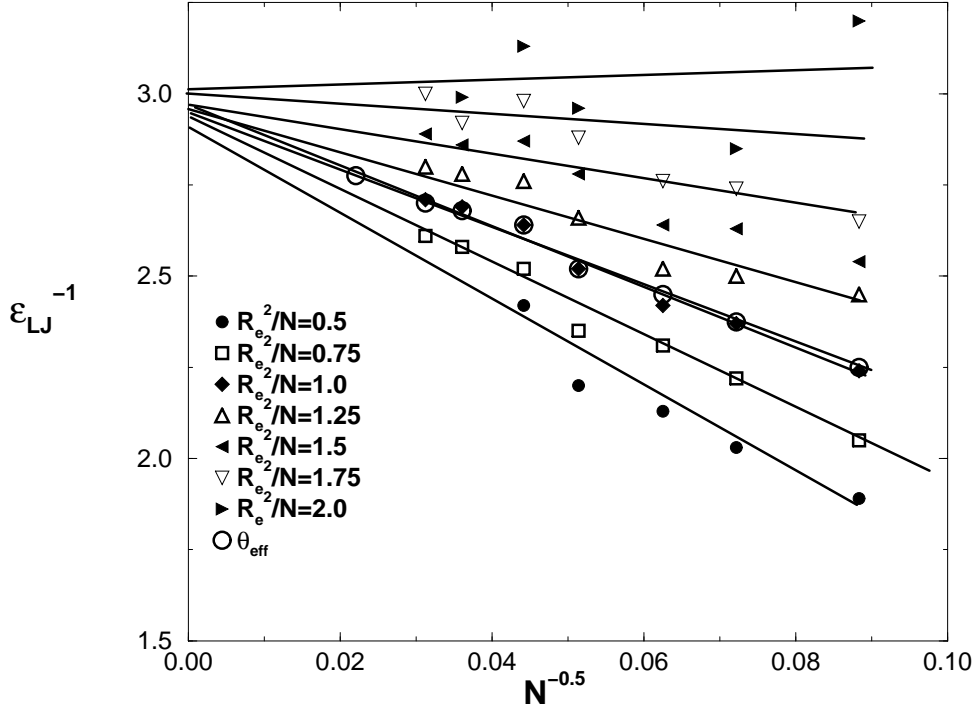,width=\textwidth}
  \end{center}
  \caption{Scaling analysis for determining the $\theta$ point of the neutral
    system. Plotted is $\epsilon_{LJ}^{-1}$ vs. $N^{-0.5}$ according to 
    Eq. \ref{neutscalequat}, where $T = \epsilon_{LJ}^{-1}$ in our units.
    One obtains straight lines
    $\text{N}^{-\frac{1}{2}} \rightarrow 0$ which converge up to corrections
    of higher order to the $\theta$ point. $\theta_{eff}$ describes the
    $\theta$ point for the finite chain, where $\epsilon_{LJ}$ was chosen such
    that its $R_{end}$ had the same value as an equally long random walk. These
    values converge as well to the $\theta$ point for $N \rightarrow
    \infty$. One obtains $\epsilon_{LJ,\theta}^{(n)} = \frac{1}{2.96\pm 0.06}
    = 0.34 \pm 0.02 $.}  
\label{thetaplot}
\end{figure}

Plotting T versus  $N^{-\frac{1}{2}}$ for fixed ratio of
$\frac{R^{2}_{end}}{N}$ results in straight lines that should converge to the
$\theta$ point for $N^{-\frac{1}{2}} \rightarrow 0$, neglecting 
some deviations due to statistical errors and higher order corrections to
scaling. Neutral chains from $N=32$ up to $N=2048$ monomers were studied. 
The scaling plot can be viewed in Figure \ref{thetaplot}, showing
that the effective interaction parameter for $\theta$ conditions is given
by $\epsilon_{LJ,\theta}^{(n)} = [2.96 \pm 0.06]^{-1} = 0.34 \pm 0.02$.

Based on this information for the neutral system three simulations of the
charged system were performed at a fixed density of 
$\rho = 10^{-4}\sigma^{-3}$, but for the three different values
$\epsilon_{LJ} = 0.25$, $0.5$, and $1.0$. 
While $\epsilon_{LJ} = 0.25$ is well
above the neutral $\theta$ point, where a behavior similar to the hydrophilic
system can be expected, $\epsilon_{LJ} = 0.5$ should show some signs
of the hydrophobic influence. For that reason, the system with  $\epsilon_{LJ}
= 0.5$ will be called ``weakly hydrophobic'' for the rest of the paper, while the
system with $\epsilon_{LJ} = 1.0$ is deep in the hydrophobic regime. 

\begin{figure}[htb]
  \begin{center}
    \leavevmode
    \epsfig{file=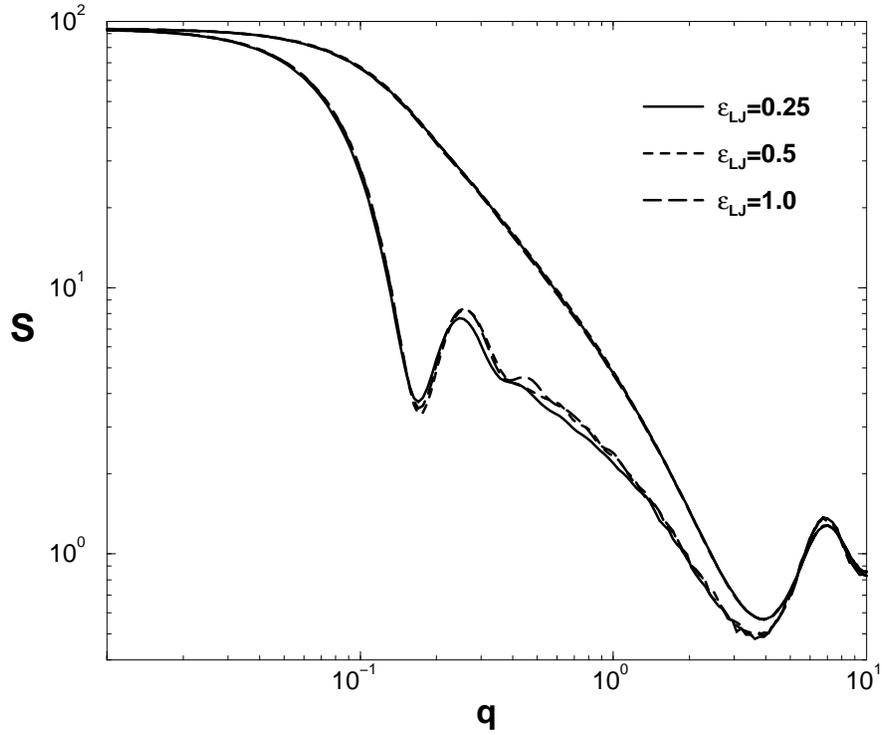, width=\textwidth}
    \caption{Comparison of the spherically averaged 
    structure factor, $S_{sp}(q)$ (top curve) and its component along the
    first principal axis of inertia, $S_\parallel (q)$ (bottom curve), versus
    $q$ for various hydrophobic 
    interaction strengths $\epsilon_{LJ} = $ 0.25, 0.5, and 1.0 at a
    density \rh{4}.} 
    \label{epsisf}
  \end{center}
\end{figure}

\begin{figure}[htb]
  \begin{center}
    \leavevmode
    \epsfig{file=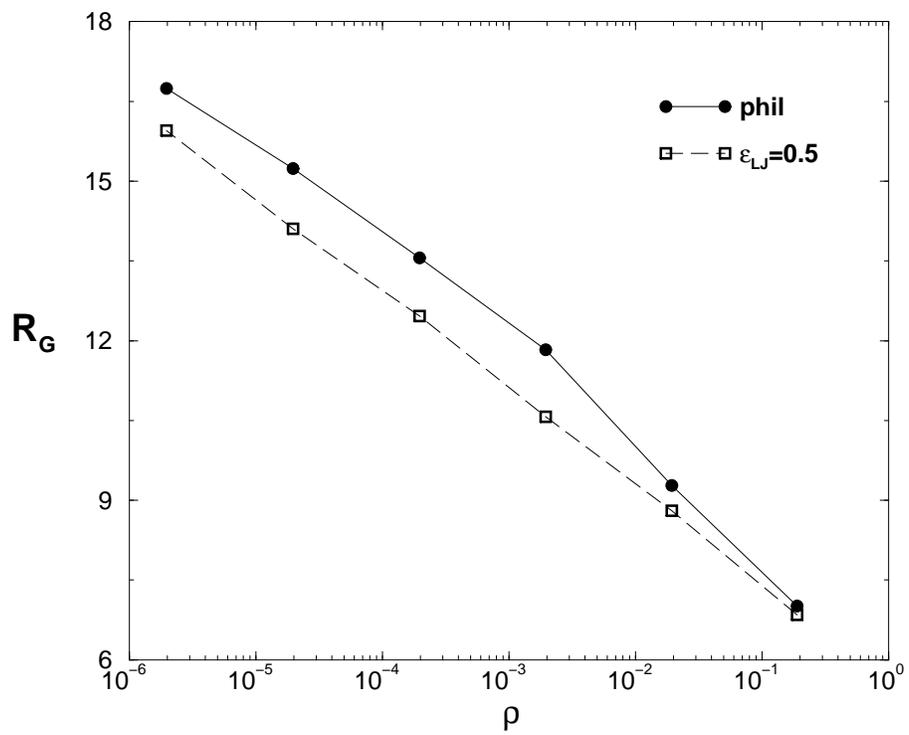, width=\textwidth}
    \caption{Comparison of the radius of gyration $R_{G}$ versus density
    $\rho$ of hydrophilic (phil) and weakly hydrophobic ($\epsilon_{LJ} =
    0.5$) polyelectrolytes.}  
    \label{rgplot}
  \end{center}
\end{figure}

\begin{figure}[htb]
  \begin{center}
    \leavevmode
    \epsfig{file=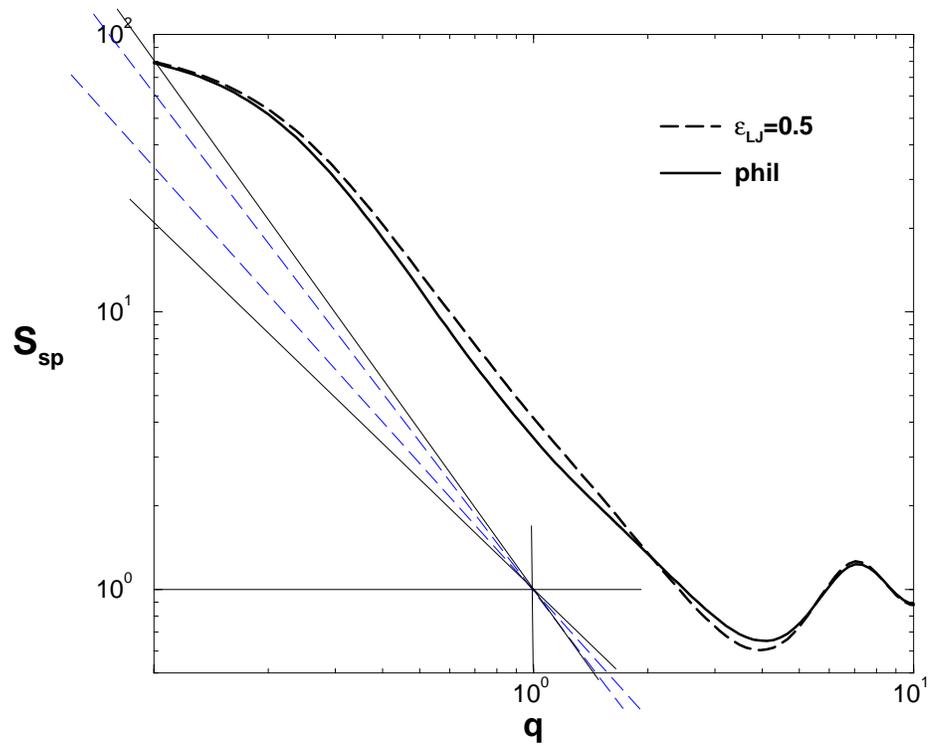, width=\textwidth} 
    \caption{The spherically averaged structure factor $S_{sp}(q)$ vs. $q$ for
    hydrophilic (phil) and weak hydrophobic ($\epsilon_{LJ} = 0.5$)
    chains for the density \rh{1}. The dashed straight lines indicate 
    the slope of the linear part of
    the hydrophobic data, the solid lines that one of the hydrophilic data.}
    \label{sfanfang}
  \end{center}
\end{figure}

We first looked at
the spherically averaged structure factor of a single chain
\begin{equation} \label{sf}
S_{sp}(q) = \left\langle \frac{1}{{\rm N}}\left|\sum_{i<j}^{{\rm N}}
\exp(i\vec q \cdot (\vec r_{i}-\vec r_{j}))\right|^2 \right\rangle
\end{equation}
and its component along the first principal axis of inertia, $S_\parallel$, 
which is obtained by restricting the $\vec{q}$ - vectors to the direction of the
calculated principal axis.
The values for $S_{sp}(q)$ and $S_\parallel (q)$
are displayed in the top and bottom part of
Figure~\ref{epsisf}, respectively, where it is found that all three of these
systems are very similar and do not differ significantly from the known
behavior of hydrophilic polyelectrolytes. 
Because the charge density strongly influences the
screening of the electrostatic interactions,
the comparison of charged systems for different values of $\epsilon_{LJ}$ is
extended over a whole range of densities. In
Figure~\ref{rgplot} the radius of gyration $R_G$ is displayed versus various
densities. The attractive potentials cause locally more compact conformations, 
leading to smaller $R_G$, which can also be seen in
Figure~\ref{sfanfang}, where the
spherically averaged structure factor is displayed for \rh{1}.

The main results of these preliminary investigations show, that the
electrostatic interaction is still too strong compared to the influence of 
the hydrophobicity. However, under certain conditions, even the ``weak''
hydrophobicity has pronounced effects.  

One of theses conditions is the relative strength of the electrostatic
interaction. The strength of the Coulomb interaction can be conveniently 
altered by varying the Bjerrum length $\lambda_{B}$. It is well known that if
one increases $\lambda_{B}$ by starting from small values, this will lead to a
non-monotonic behavior of the chain dimensions \cite{Mark1}, which is due to
counterion condensation\cite{Manning,And,Oos}. 

The basic idea of counterion condensation is a generalization of the well
known solution of Onsager \cite{Onsager} for the infinitely long and thin line
charge. If the energy density on the rod exceeds a certain value it becomes
more favorable to catch one of the free ions to neutralize one of the charges
on the chain. The calculation results in the condition that the distance
between two adjacent unit charges on the chain has to be bigger than
$\lambda_{B}$, 
which means that their mutual pure Coulomb repulsion energy has to be smaller
than $k_{B}T$. For that reason, the entropy of the counterion, that is lost in
the condensation process, is assumed to equal $k_{B}T$. By analogy, this
argument is applied to any (flexible) polyelectrolyte, which is, however,
questionable. 
We define a counterion as condensed, if its total energy is smaller
than $-1 k_{B}T$. This way is easy to handle and
strongly supported by a visual inspection of the set of condensed counterions 
of our conformations. It has some advantages over the geometrical definition,
where one looks at the distance of a counterion to its nearest polymer, and
sets some arbitrary bound, because an ion can feel a strong attraction to
several charged monomers even at a relatively large distance.

\begin{figure}[htb]
  \begin{center}
    \leavevmode
    \epsfig{file=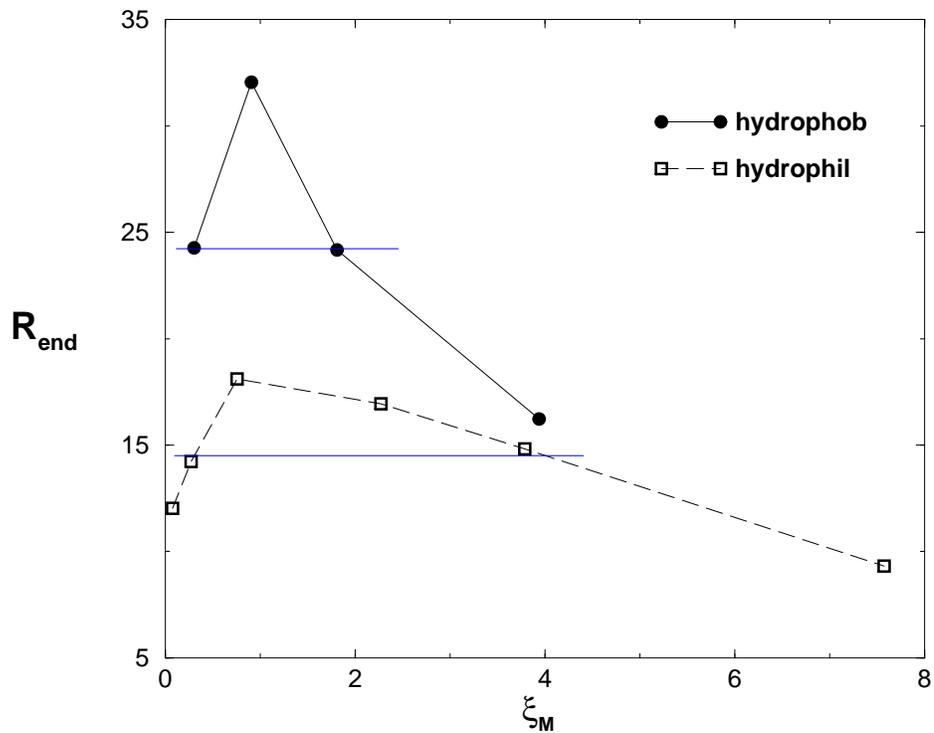, width=\textwidth}
    \caption{End-to-end distance $R_{end}$ as a function of the Manning
    parameter $\xi_M$ for hydrophilic chains ($\epsilon_{LJ} = 0.25$, $\Box$,
    from Ref. \cite{Mark1}), and hydrophobic chains ($\epsilon_{LJ} = 
    0.5$, $\bullet$). The  
    straight line indicates comparable end-to-end distances.}
    \label{rxsi}
  \end{center}
\end{figure}

The results of Ref.\cite{Mark1} and our new data 
show that first, increasing $\lambda_B$ stretches the
chains through the increased repulsion of the equally charged chain monomers. 
Then a maximum is reached and the chains start to contract, due to counterion
condensation, which decreases the effective  
charge on the chain. When the Bjerrum length is further increased, eventually
all counterions condense, and then the collapsed chains will start to
coagulate forming a liquid sol and droplet. 
This was later also observed in Ref. \cite{winkler97}.
Because the hydrophobic polyelectrolytes have an additional attractive force
between the monomers, a reduction of the chain extensions occurs already 
for lower $\lambda_{B}$, and this contraction will be
stronger than in the hydrophilic case. Figure \ref{rxsi} reflects
exactly this behavior, where we compare the end-to-end distances $R_{end}$ 
for the two cases $\epsilon_{LJ} = 0.25$ and $\epsilon_{LJ} = 0.5$ as
a function of the Manning ratio $\xi_M$ at a density of \ro{2}. 
The data for the hydrophilic
polyelectrolytes is taken from Ref. \cite{Mark1}. where the model
contained no explicit neutral monomers, so we can not 
compare the absolute values. For our purposes, however, a comparison of the
tendencies is sufficient. In Figure \ref{rxsi} one recognizes that the
hydrophobic system contracts to a radius that is a factor of two smaller at
$\xi_M = 3.94$! The relative contraction of the hydrophilic chain compared to
the reference state is only about 30\% at $\xi_M =  7.58$. These values
clearly show that already the influence of weak hydrophobicity changes
drastically the conformational properties of polyelectrolytes. 

\begin{figure}[htb]
  \begin{center}
    \leavevmode
    \epsfig{file=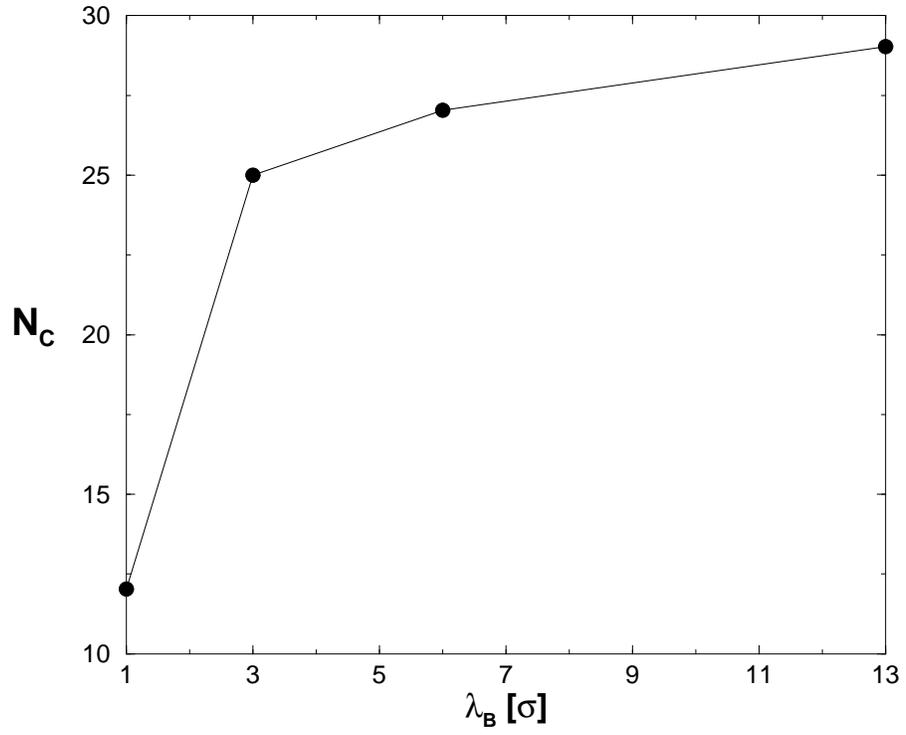, width=\textwidth}
    \caption{Number of condensed counter ions $N_C$ as a function of the
    Bjerrum length \lb for $\epsilon_{LJ} = 0.5$ at the density \ro{3}. The
    number of counterions per chain is 32.}
    \label{lbplot}
  \end{center}
\end{figure}

\begin{figure}[htb]
  \begin{center}
    \leavevmode
    \epsfig{file=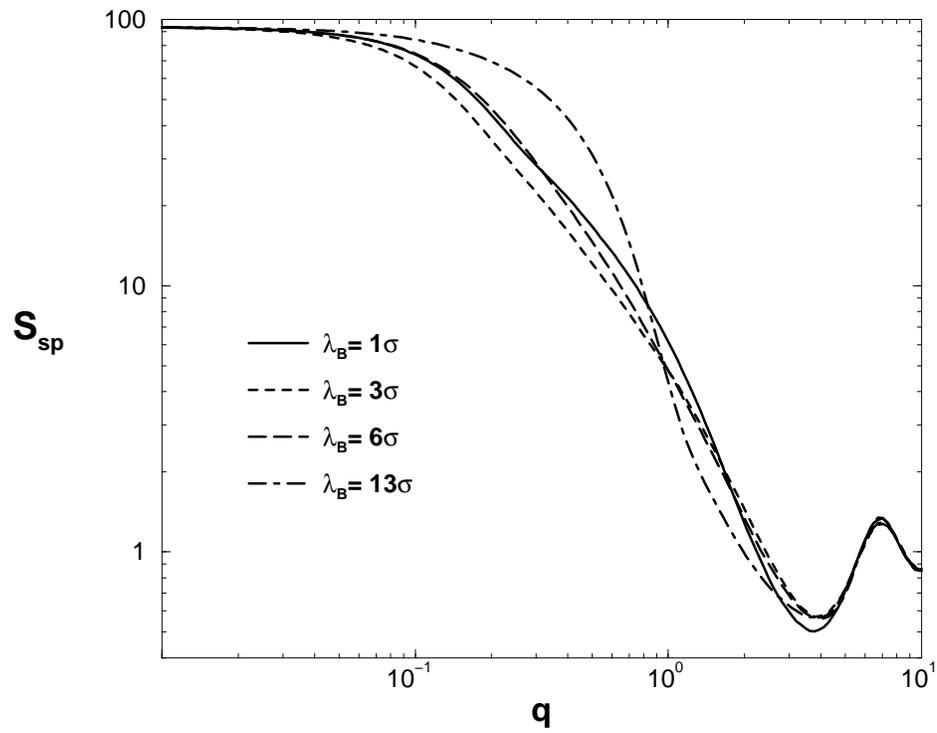, width=\textwidth}
    \caption{Spherically averaged structure factor $S_{sp}(q)$ vs. $q$ for
    various Bjerrum lengths \lb for hydrophobic chains with $\epsilon_{LJ} =
    0.5$.} 
    \label{2sflb}
\end{center}
\end{figure}

Further evidence for this can be
gained from the number of condensed
counterions as a function of the Bjerrum length $\lambda_B$, displayed in
Figure~\ref{lbplot}. It increases from 12 to 29 around $\lambda _B \approx 3$
($\xi_M \approx 1$) showing exactly the dramatic effect
as argued above and in reference \cite{Mark1}. An analysis of the structure
factor at the very high Bjerrum length of $\lambda_{B} = 13\sigma$, compare
Figure \ref{2sflb}, gives
further evidence that the conformation is in a collapsed state.

\begin{figure}[htb]
  \begin{center}
    \leavevmode
    \epsfig{file=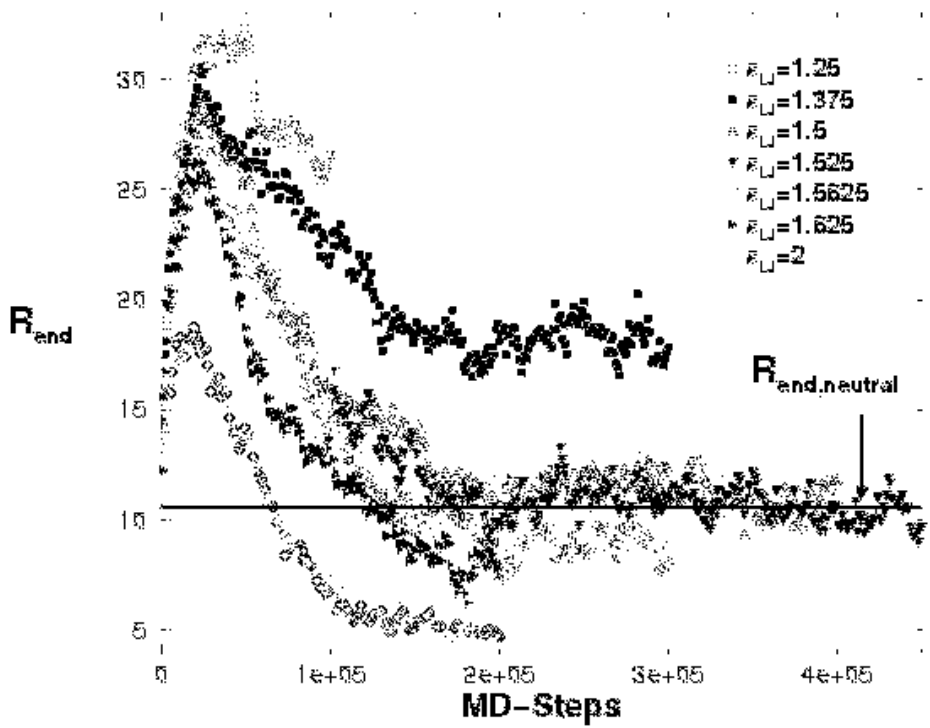, width=\textwidth}
    \caption{Time development in MD steps of the end-to-end distance $R_{end}$
    of charged chains for various hydrophobicities $\epsilon_{LJ}$. The
    straight horizontal line indicates the end-to-end distance of the equally
    long uncharged random walk}
    \label{rendzeitreihe}
  \end{center}
\end{figure}

As seen above, even at a value of
$\epsilon_{LJ} = 1.0$ the chain conformation at \rh{4} was not 
significantly altered although the neutral chain would be strongly in the
hydrophobic regime. For that reason, we found it necessary to 
compute an ``effective'' $\theta$ point for the charged system 
in the following way:
The end-to-end distance $R_{end}$ is calculated as a function of
$\epsilon_{LJ}$. The effective $\theta$ point is then defined by the value of
$\epsilon_{LJ,\theta_{eff}}^{(c)}$ at which $R_{end}$ equals the one for an
ideal 
neutral chain of this length. The $\theta$ point is called effective, because
it  
is determined only for the system under consideration, and hence is not an
asymptotic statement. 
Figure \ref{rendzeitreihe} shows the time development of $R_{end}$ for
different $\epsilon_{LJ}$ at the fixed density \rh{4}. 
From the data the relevant parameter $\epsilon_{LJ,\theta_{eff}}^{(c)}$ is
determined to be 
\begin{equation}
\epsilon_{LJ,\theta_{eff}}^{(c)} \simeq 1.51 \pm 0.03.
\end{equation}
Polymers with this parameter will be henceforth called ``strongly
hydrophobic''.
Starting from this value of $\epsilon_{LJ,\theta_{eff}}^{(c)}$, dominating
hydrophobicity at higher densities and dominating electrostatics at lower
densities will be observed in the next section. 
%
\section{Variation of the density - single chain properties \label{konf}}
%
Our intention in this section is to study the crossover from strong
screening to dominating electrostatic interaction by changing 
the polymer density $\rho$. The Bjerrum length is fixed to
$\lambda_{B}=3.0\sigma$, $k_{B}T = 1$ and the strength of the hydrophobicity is
set to $\epsilon_{LJ} = 1.5$ in light of the results of the previous section. 
16 chains of length $N=94$ are simulated for densities varying from
\ro{1} to \ro{6}. 

\begin{figure}[htb]
  \begin{center}
    \leavevmode
    \epsfig{file=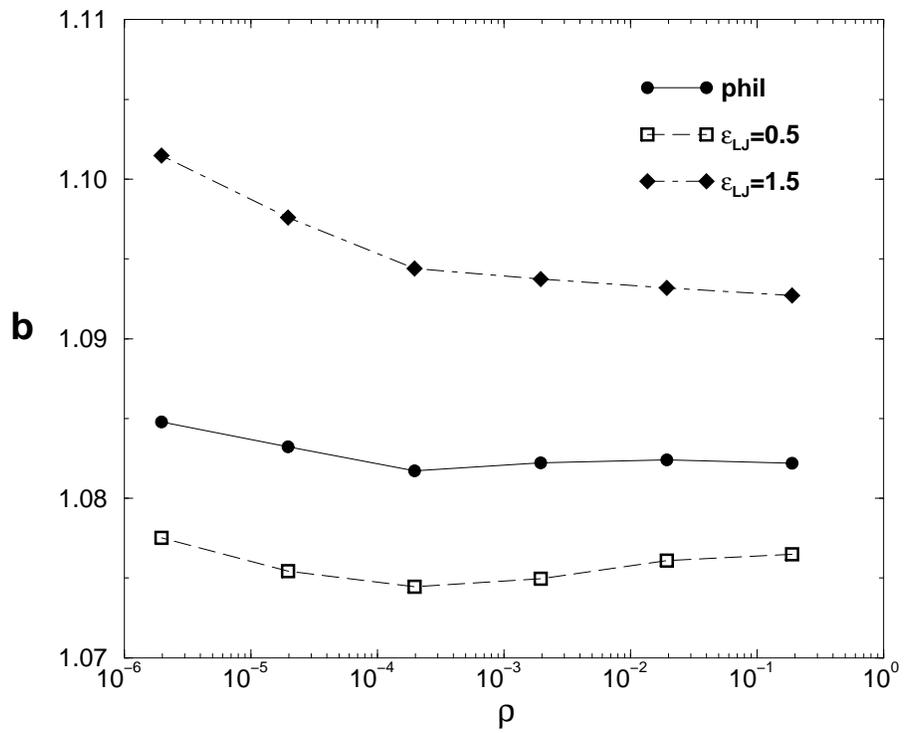, width=\textwidth}
    \caption{Bond length $b$ versus density $\rho$ for hydrophilic (phil), 
      weakly
    hydrophobic ($\epsilon_{LJ} = 0.5$), and hydrophobic ($\epsilon_{LJ} =
    1.5$) polyelectrolyte chains.} 
    \label{bind}
  \end{center}
\end{figure}

First we want to start by comparing a
hydrophilic system to the weakly hydrophobic one,
$\epsilon_{LJ} = 0.5$, and to the strongly hydrophobic system with 
$\epsilon_{LJ} = 1.5$.
Especially noteworthy is the similarity between 
the weakly hydrophobic and the hydrophilic system, and the strong contrast of
both systems to the strongly hydrophobic system.
On a very local scale there is no large difference in all three systems 
as a function of density.
This is reflected in Figure \ref{bind} that shows the average bond length. The
differences in the total values are due to the slight differences in the
repulsive part of the LJ potentials with different $\epsilon_{LJ}$. As 
this difference is only small and completely irrelevant for the rest of the
investigation, we did not 
correct for it, e.g. by an adjustment of the FENE potential. 

\begin{figure}[htb]
  \begin{center}
    \leavevmode
    \epsfig{file=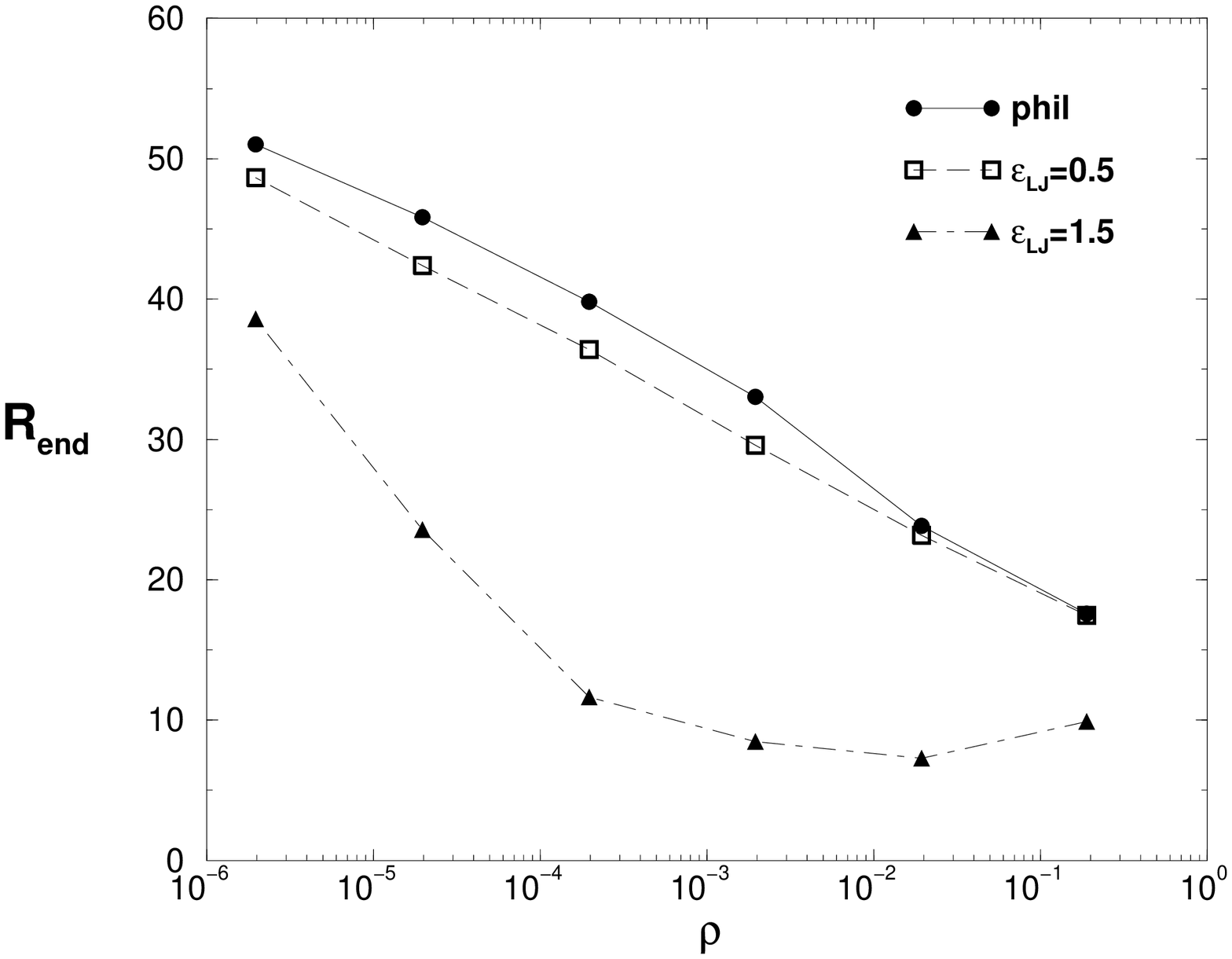, width=10 cm}
    \epsfig{file=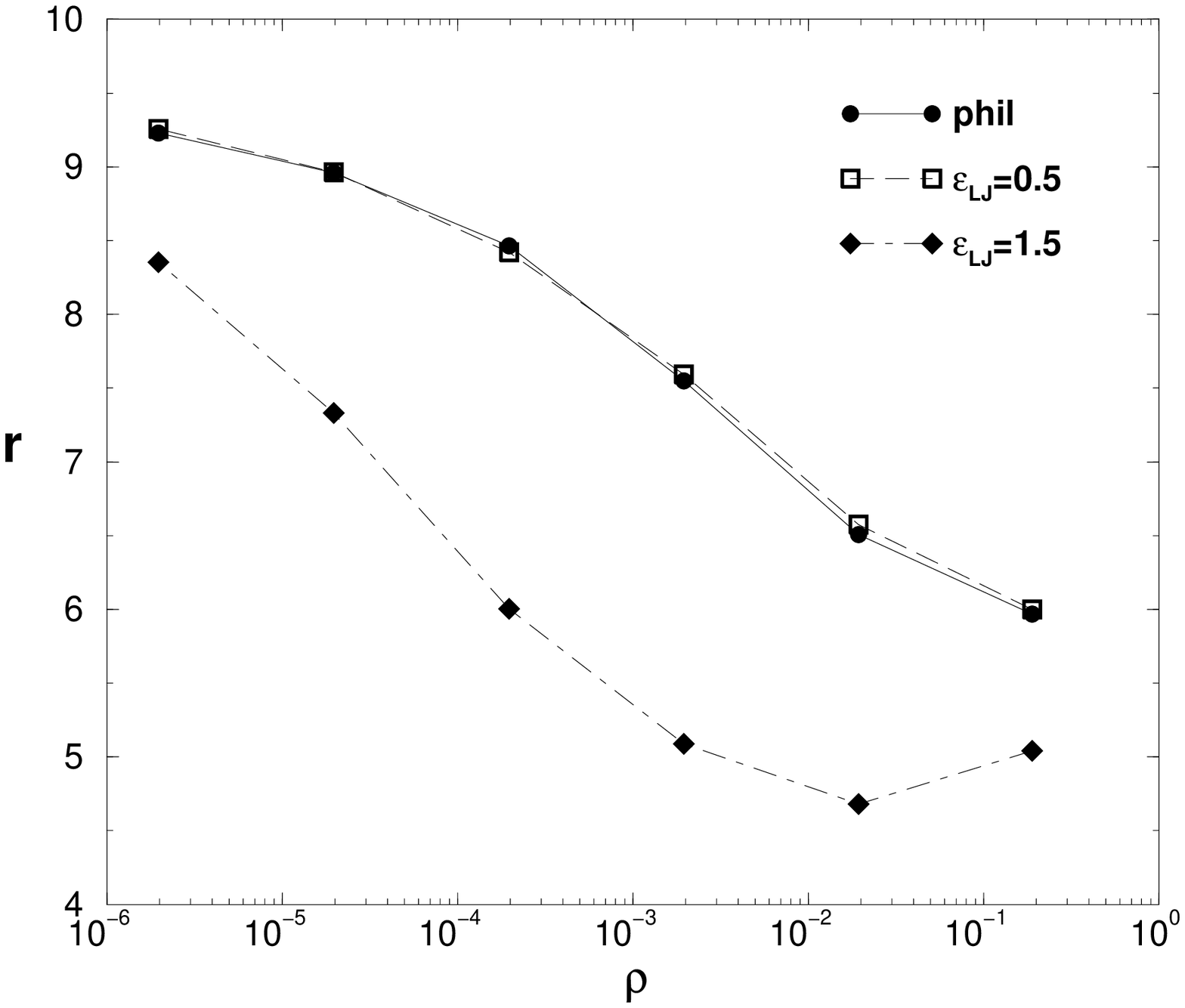, width=10 cm}
    \caption{Density dependence of the end-to-end distance $R_{end}$ (top)
    and the statistical ratio $r$ (bottom) vs. density $\rho$ for hydrophilic  
    (phil), weakly hydrophobic ($\epsilon_{LJ}=0.5$), and strongly hydrophobic
    ($\epsilon_{LJ}=1.5$) systems. }
\label{prend}
  \end{center}
\end{figure}

\begin{figure}[htb]
\vspace*{-5cm}
  \begin{center}
    \leavevmode \vspace*{-2cm}
    \epsfig{file=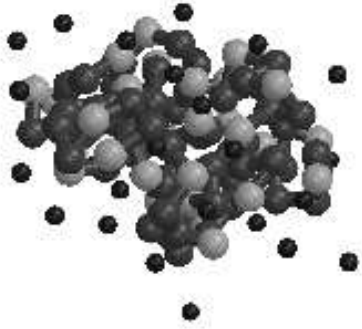, width=9 cm}
    \epsfig{file=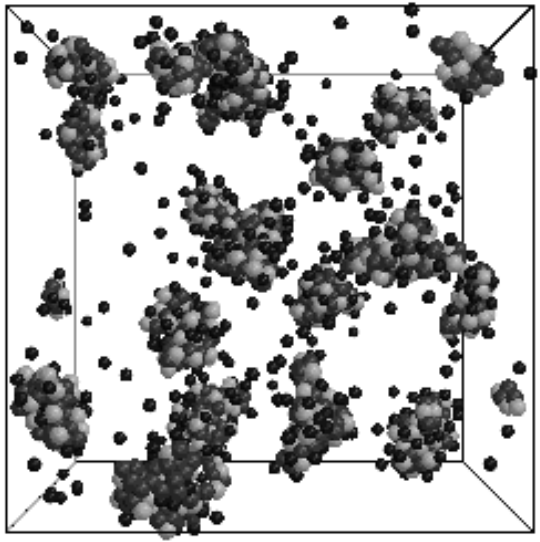, width=9 cm}
    \caption{Conformations at the density \ro{2}.
      Top: A typical poor solvent polyelectrolyte conformation. The beads are
    the charged monomers, the counterions are 
    indicated as small black spheres, and the neutral monomers are dark grey
    spheres. Only counterions within a distance of $3\sigma$ to the chain are
    displayed.
    Bottom:
    A snapshot of the whole simulation box, showing all 16 chains
    together with their counterions. The picture shows that the chains 
    collapse into single globules which are well separated, and a small
    fraction of the counterions is still in solution.}  
    \label{poly2}
  \end{center}
\end{figure}

Much more interesting is
the behavior of the end-to-end distance $R_{end}$ and the characteristic
ratio $r = \frac{R^{2}_{end}}{R^{2}_{G}}$.
As already presented elsewhere \cite{letter} the
hydrophilic system stretches for decreasing density from a starting point well
above the self-avoiding-walk value. The hydrophobic chains pass through a
pronounced minimum at \ro{2}. Both observables can be inspected as a function
of density in Figure \ref{prend}. The characteristic
ratio $r$ is known to equal six for a
random walk, 12 for a rod and is supposed to be around 6.3 for a self avoiding
walk. For a homogeneous sphere, with the ends randomly distributed over the
volume, $r = 2$, and $r = 10/3$, if the ends are constraint to be randomly
distributed on the surface \cite{duenweg98,letter}. 
For deformed collapsed chains we
thus expect a value well below $r=6$ but above $r = 10/3$. For
the density \ro{2} we find a value of $r=4.68$, meaning that the
system is in a strongly collapsed state. 
A typical conformation for that density and the snapshot of the whole box 
can be viewed in Figure \ref{poly2}. 

\begin{figure}[htb]
  \begin{center}
    \leavevmode
    \epsfig{file=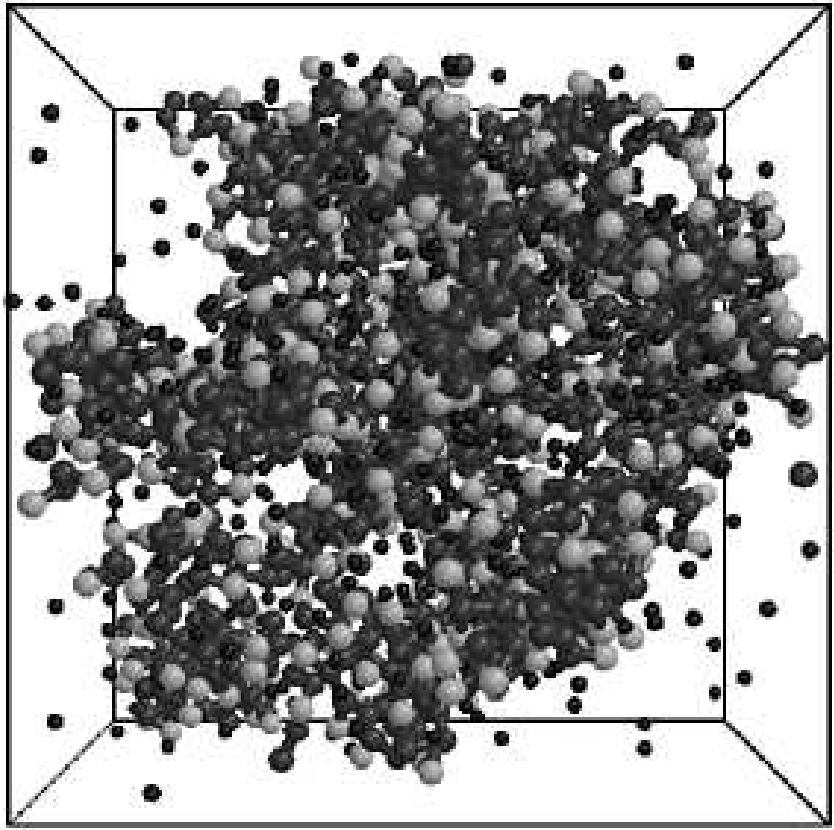, width=\textwidth}
    \caption{A snapshot of the whole simulation box, showing 16 chains
    together with their counterions for the density \ro{1}. The beads are the
    charged monomers, the counterions are indicated as small black spheres,
    and the neutral monomers are dark grey spheres.}
    \label{konf1}
  \end{center}
\end{figure}

\begin{figure}[htb]
  \begin{center}
    \leavevmode
    \epsfig{file=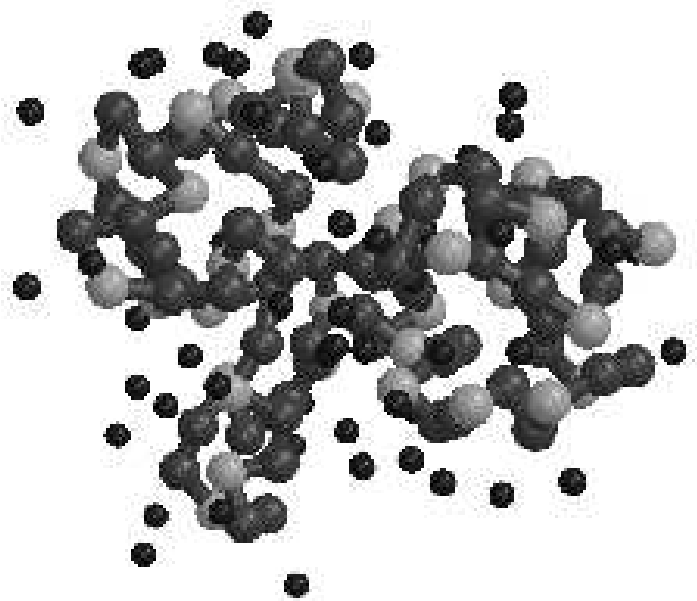, width=\textwidth}
    \caption{A typical poor solvent polyelectrolyte conformation for the
    density \ro{1}. The coloring is the same as in Fig.\ref{poly2}} 
    \label{poly1}
  \end{center}
\end{figure}

\begin{figure}[htb]
  \begin{center}
    \leavevmode
    \epsfig{file=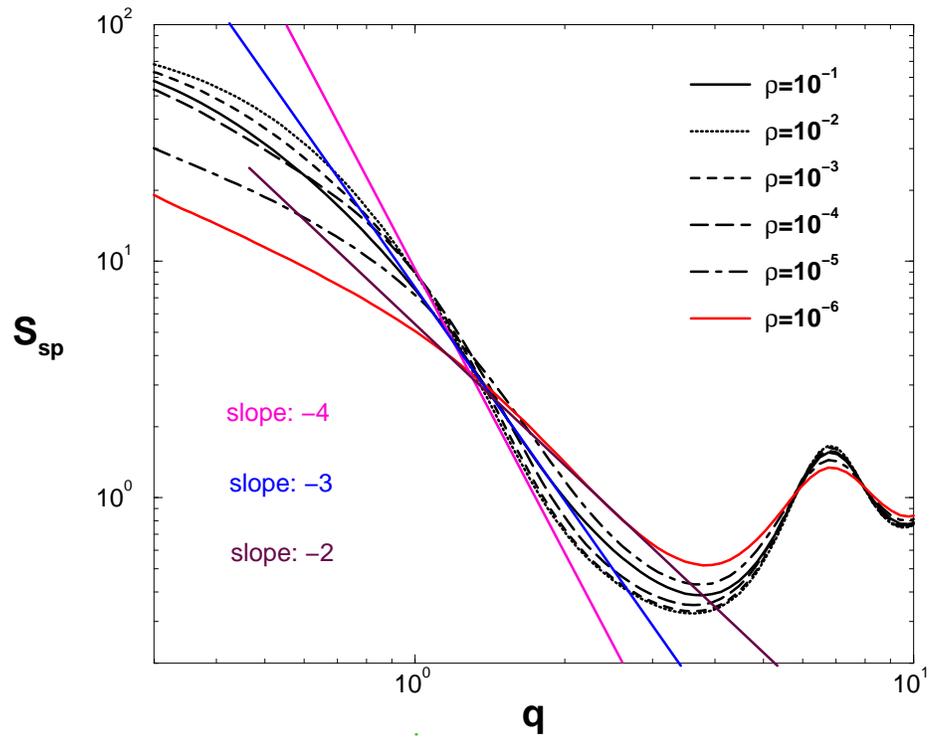, width=\textwidth}
    \caption{Details of the spherically averaged structure factor $S_{sp}(q)$ for
    large $q$ for the indicated densities $\rho$. The straight lines have
    slope $m=-2, -3$, and $-4$, respectively.}  
    \label{sfklein}
  \end{center}
\end{figure}

At the highest density $\rho = 0.2 \sigma^{-3}$, where the electrostatic
interaction is
even weaker, the chains are not collapsed completely. However, the
density is so high that strong intermolecular hydrophobic forces cause the
system to collapse into a gel-like object, as can be inspected in Figure 
\ref{konf1}, where every chain is interpenetrated by several others.
The conformations of the 
single chains are dominated by the topological constraints imposed by
neighboring chains, and we find a broad spectrum of shapes. 
A simple screening argument for the electrostatic and the excluded volume
interaction can not explain the increase in size, because the density is still
too low. However, there might also be problems connected with the simulation
of a quasi-stable (eventually also in experiment ``glassy'') state. For these
higher densities, there is almost no energetic advantage to collapse in well
separated single molecules any more.
Figure \ref{poly1} shows a typical representative of the dominant
conformations. It is clearly less collapsed than the chain at the lower
density in Figure \ref{poly2}. 
This can also be observed more clearly in the single chain structure factor. A
collapsed chain should have a 
chain length dependence of $R_{end}$ that scales with an exponent $\nu =
\frac{1}{3}$ which leads to a decay of the spherically averaged structure
factor 
with a slope of $m = -3$. This is roughly seen in the data set at
$\rho= 0.2 \sigma^{-3}$. For \ro{2}, however, a value of $m=-4$ is found
that is characteristic for Porod scattering at distinct phase boundaries
\cite{Porod}. This demonstrates how extreme the chain collapse is, as can be
inspected  
in Figure \ref{sfklein}. For lower densities the slope converges towards
$m = -2$, which is the random walk value. Nevertheless, this value has to be
taken with care, because the attractive forces will prevent one
from observing a real random walk behavior. Additionally, the
linear range in the log-log plot of Figure \ref{sfklein} is rather small.

We have found that the collapse of the system 
into a microgel, or isolated chains, is just a consequence of a strongly 
hydrophobic attraction in polyelectrolytes.
Because the simulated densities were rather high, this effect should be easily
observed in experiments.
Especially the isolated chains should be a very interesting candidate 
to study, because they still interact strongly via
electrostatic interactions. As Figure \ref{poly2} shows, they should behave
like a solution of charge stabilized colloidal particles.
Nevertheless, the internal
degrees of freedom will always play a very important role in polyelectrolytes
and distinguish them from charged colloids. 

\begin{figure}[htb]
  \begin{center}
    \leavevmode
    \epsfig{file=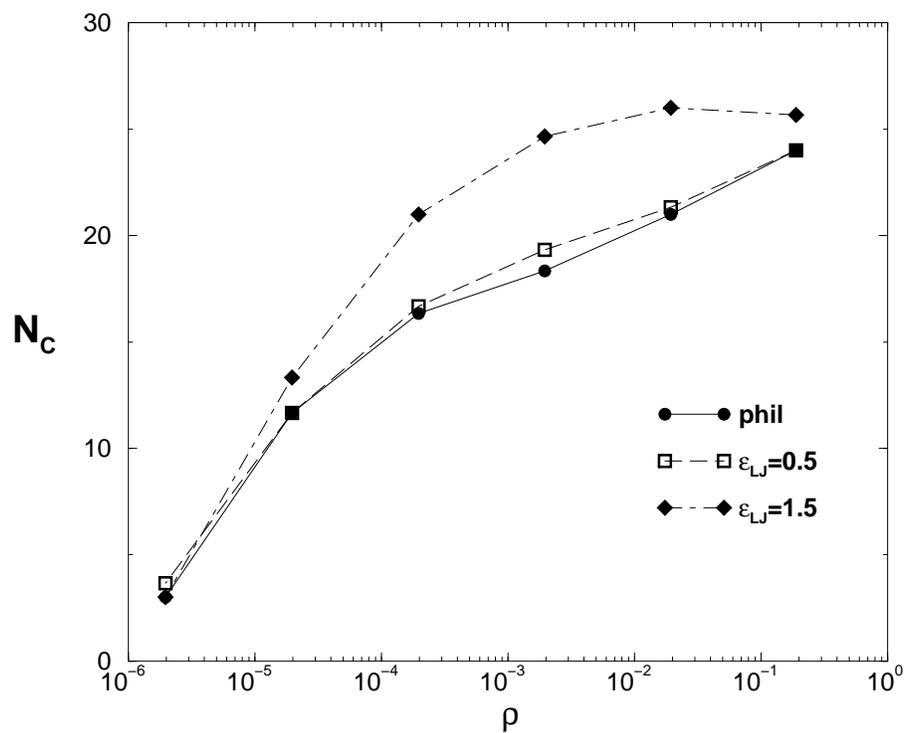, width=\textwidth}
    \caption{The number $N_c$ of condensed counter ions per chain as a
    function of density $\rho$ for $32$ charged monomers per chain for
    hydrophilic chains (phil) and hydrophobic chains with $\epsilon_{LJ} =
    0.5$ and 1.5.} 
    \label{ccplot}
  \end{center}
\end{figure}

The stability of the colloidal
phase can be inferred by the following simple estimates\cite{letter}.
We start to determine the number of condensed ions in order to get the net
charge of the collapsed object. The results are plotted in Figure
\ref{ccplot}. We find that
26 ions are condensed at \ro{2} leading to a net charge of 6 per chain.
This coincides very nicely with the number which one obtains from
the critical charge $Q$ for a charged drop according to the Rayleigh
criterion, where one balances the Coulomb energy 
$E^{Coul}(R_G) = \lambda_B k_B T Q^2/ R_G$
with its surface energy
$E^{surf} (R_G) = 4 \pi R_G^2 k_B T/A_{surf}$. If
one uses for the surface area $A_{surf}$ 
the monomer area $\pi b^2$, one gets for the
critical charge the formula
\begin{equation}
Q^2 = \frac{4R_G^3}{b^2\lambda_B}.
\label{rayleigh}
\end{equation}
Using the values $b = 1.08\sigma$ and $R_G = 3.2 \sigma$, valid for the
density \ro{2}, one finds $Q \approx 6.1$.
Of course the number of condensed counterions strongly increases with density,
because the screened electrostatic interaction cannot stretch the chains any
longer. These more compact states have to be stabilized by the oppositely
charged counterions. Their entropy loss is overcome by the energy gain in the
chain.
In Ref.~\cite{letter} a simple free energy argument was presented to show that
the colloidal phase (1c) is more stable than a hypothetical agglomerate of two
chains (2c). The critical charge $Q_{2c}$ for the agglomerate of two chains
for the same density \ro{2} is found with Eq.(\ref{rayleigh}) by using 
$(R_G)_{2c} = R_G2^{1/3}= 4.0 \sigma$ to take on the value $Q_{2c} \approx
8.7$. Therefore, in order for two chains to agglomerate an 
additional number of 3.5 counterions have to
condense on average onto the pair of chains which carry each an average charge
of $Q_{c1} = 6.1$. To estimate the free energy difference of 
the two globular states, one
needs to look only at the Coulomb energy of the globules with each other and
the entropy of the remaining counterions in solution, because the intra
Coulomb energy and the surface energy are already balanced from the Rayleigh
criterion (\ref{rayleigh}). 
Assuming a uniform distribution of the chains in the simulation box, we find
the distances $d_{1c} = 18.5 \sigma$ and $d_{2c} = 23.3 \sigma$ for the
one chain and two chain globular state, respectively, to their
next neighbors. 
The repulsive Coulomb energy between two neighboring chains is given by
\begin{equation}
E^{Coul} = \lambda_{B} k_{B}T \frac{Q^2}{d}.
\end{equation}

To make the calculation simple we assume the chains to sit on a simple cubic
lattice. We also neglect higher multi-pole moments and take only the next
neighbor chains into account. The influence of the counterions is also
ignored, because on the average 
only one is available per neighbor. With these simplifications
the average Coulomb interaction energy per single chain is found to be
$E^{Coul}_{1c} = 18.1 k_{B}T$/chain and $E^{Coul}_{2c} = 14.6 k_{B}T$/chain,
respectively. 
The corresponding value for an agglomerate of two chains is therefore about
$3.5k_{B}T$ lower, preferring the agglomerate.
The entropy of the free counterions can be approximated by an ideal gas
Ansatz. The free counterion densities are given by $\rho_{1c} = 6.1/(94+32)
\rho$ and $\rho_{2c} = 4.35/(94+32)\rho$, respectively.
Because the entropy per free ion is given by the logarithm of the available
volume fraction the entropic contribution of the free energy difference per
single chain is
given by $6.1 \ln \rho_{1c} - 4.35 \ln \rho_{2c} \approx -10.7 k_B T$. This
gives a total preference of $7.2 k_B T$/chain for the single chain
globule\cite{letter}.  

\begin{figure}[htb]
  \begin{center}
    \leavevmode
    \epsfig{file=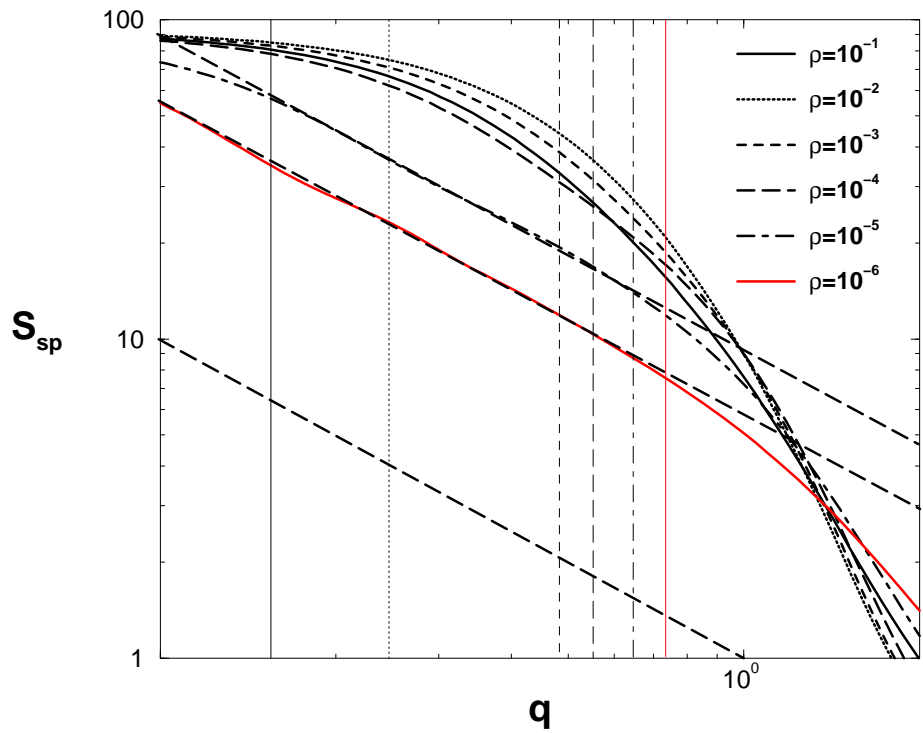, width=\textwidth}
    \caption{Details of the spherically averaged structure factor $S_{sp}(q)$
    vs. $q$ for
    small $q$, obtained for the indicated densities $\rho$. The straight lines
    have all slope $m=-1$, and the vertical lines denote the value of $q =
    2\pi/R_G$.}  
    \label{sfgross}
  \end{center}
\end{figure}

\begin{figure}[htb]
  \begin{center}
    \leavevmode
    \epsfig{file=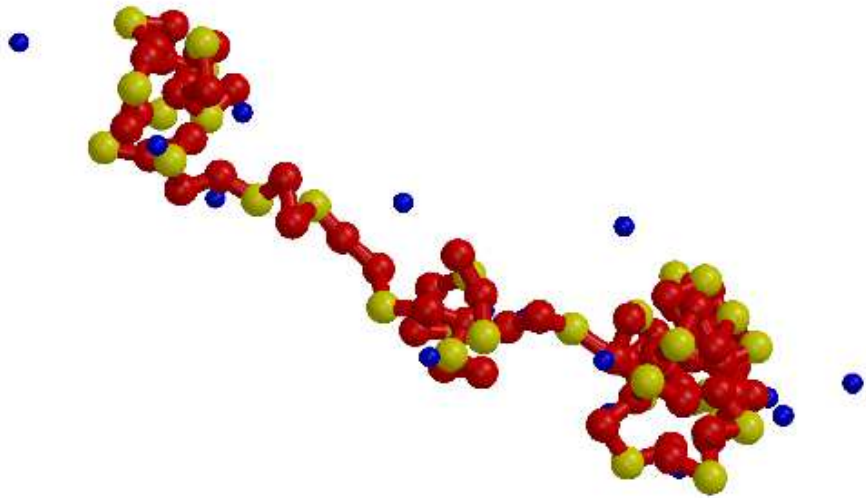, width=\textwidth}
    \caption{A typical poor solvent polyelectrolyte conformation for the
    density \ro{5}, which shows nicely a pearl-necklace structure. The
    coloring is as in Fig.~\ref{poly2}} 
    \label{poly5}
  \end{center}
\end{figure}

\begin{figure}[htb]
  \begin{center}
    \leavevmode
    \epsfig{file=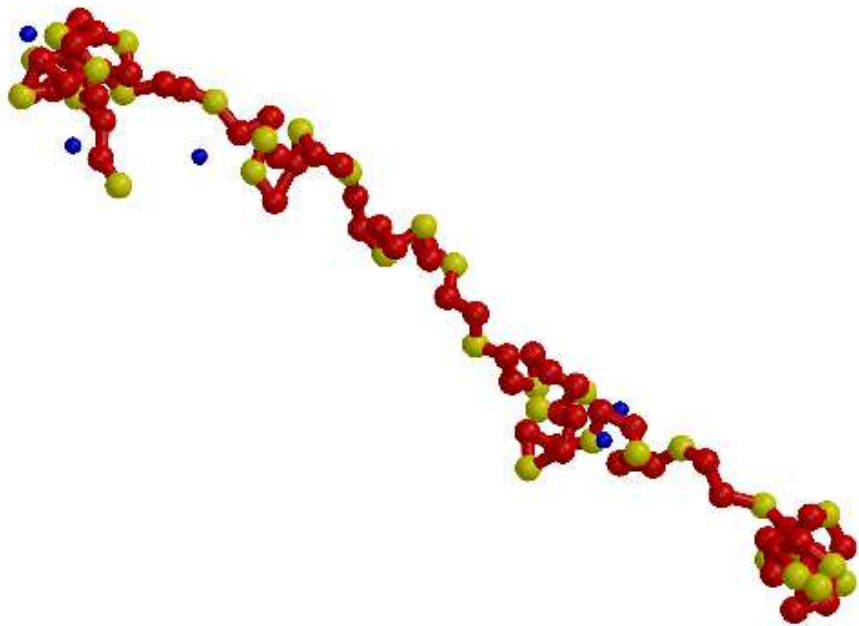, width=\textwidth}
    \caption{A typical poor solvent polyelectrolyte conformation for the
    density \ro{6}, showing a very elongated, but still visible pearl-necklace
    structure. The coloring is as in Fig.~\ref{poly2}}
    \label{poly6}
  \end{center}
\end{figure}

\begin{figure}[htb]
  \begin{center}
    \leavevmode
    \epsfig{file=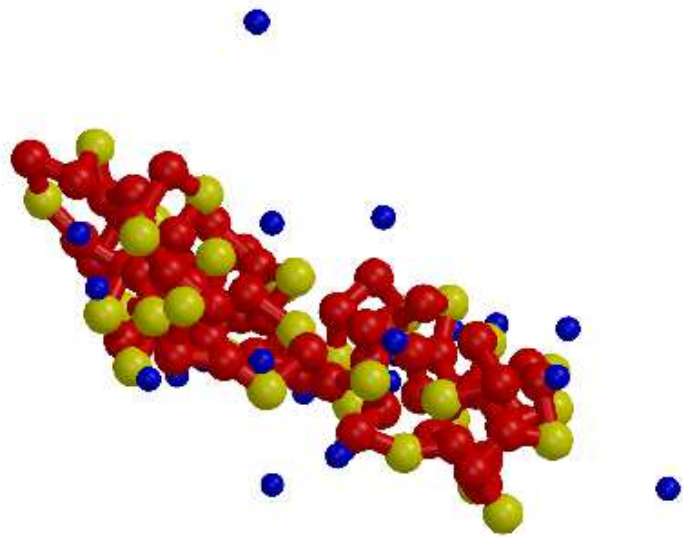, width=\textwidth}
    \caption{A typical poor solvent polyelectrolyte conformation for the
    density \ro{3}. One sees the onset of the Rayleigh instability, the
    globule wants to split up into two parts. The coloring is as in
    Fig.~\ref{poly2}.} 
    \label{poly3}
  \end{center}
\end{figure}

\begin{figure}[htb]
  \begin{center}
    \leavevmode
    \epsfig{file=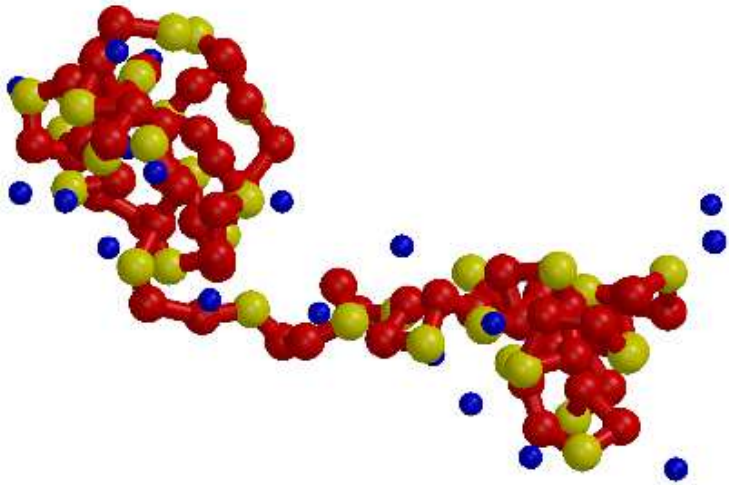, width=\textwidth}
    \caption{A typical poor solvent polyelectrolyte conformation for the
    density \ro{4}, where now two globules with a bridge connecting them, has
    formed . The coloring is as in Fig.~\ref{poly2}}
    \label{poly4}
  \end{center}
\end{figure}

The spherically averaged single chain structure factor shows another very
striking feature at low $q$, this time for the lowest densities. At \ro{5} and
\ro{6} a second well
defined length scale shows up. The logarithmic slope measured is exactly
$m=-1$, as can be seen in Figure \ref{sfgross}, which corresponds to an
exponent $\nu = 1$. This is an extraordinary result, because no other 
polymeric systems show such a behavior. For example, hydrophilic
polyelectrolytes always have an exponent $\nu < 1$, because entropic
fluctuations lead to some local roughness\cite{Mark1}. Even the
blob pole conformations observed in simulations of single Debye-H\"uckel chains
\cite{paper1} that show a linear dependency of the end-to-end distance on the
contour length after introduction of the concept of electrostatic blobs, show
an effective exponent $\nu$ smaller than one.
Figure \ref{poly5} and \ref{poly6} show typical pearl-necklace 
like conformations.
Small globules are connected by thin bridges. The stronger the electrostatic
interactions are, the larger is the number of globules. The globules are in
addition stabilized by a few condensed counterions. So we can qualitatively
reproduce the picture of Ref.~\cite{neckl} for our strongly charged systems.
Of course, the globules are much smaller than in their weakly charged case, but
clearly detectable. The chains are strongly stretched due to the electrostatic
interaction which orients the different globules to an almost rodlike object by
maximizing their mutual distance. Additionally, the chains are, at least
within the 
globules, locally not very flexible, because the monomers dislike to break the
energetically favorable hydrophobic bonds (around $E=1.5 k_{B}T$ per contact).
To complete the set of pictures of conformations, we display in
Figures \ref{poly3} and \ref{poly4} typical
conformations at intermediate densities. A continuous  deformation  towards
a dumbbell like conformation is observed, but no cylindrical instability 
is found. It is interesting to note, that the larger globules are found at the
chain ends, and the smaller ones reside in the chain middle.

At the end of this section we want to stress again, that no micro-phase
separation is observed, although this is theoretically predicted
\cite{KoKa}. Only for the highest density we can not rule out that this will
eventually happen. The main argument for a micro-phase separation was
that the entropy of the condensed counterions in the 
polyelectrolyte-rich phase is much larger, because the condensed counterions
can explore the whole volume of the separated phase. We have been showing in
our previous 
communication \cite{letter}, that the system prefers single chain globules,
which repel each other because they still have a sufficiently high net charge.
Additional counterions would need to condense to
stabilize an agglomerate of two or more chains. However, their loss of
translational entropy (and the increase in Coulomb energy) is much larger than
the gain in surface energy, and therefore single, isolated globules are
clearly favored in certain density regimes.
%
\section{Comparison with experiments \label{exp} }
%
It is important to try to compare these results to experimental
data. Unfortunately, 
most experiments do not examine single chain 
properties with adequate precision.
Especially for low densities, where the mutual influence of the chains can be
neglected, scattering
experiments are extremely difficult. A comparably new experimental study
\cite{willi2} shows good agreement with our data at relatively high
densities. Unfortunately, the density was only varied once
by a factor of two, so that the effects we report in this paper, 
could not be seen. 

Most experimental results come from detailed measurements of the structure
factor of the whole sample or thermodynamic properties like the
osmotic pressure $\Pi$ are measured. 
To compare our results of section \ref{konf}, however, 
data of single chain properties are needed, which unfortunately are rare.
The structure
factor of the whole simulation box that reflects the arrangement of the chains
in space, is difficult to determine from the simulation data for the present
poor solvent situation. 
The reasons are of 
technical nature. The structure factor itself can easily be calculated,
however it is strongly influenced by
finite size effects, as the number of chains in the simulation box 
is relatively  small (16 or 32).
The collapsed chains can see each other only weakly, which results in broad
maxima in the pair-correlation function and in $S(q)$, which was not the case
for the hydrophilic systems \cite{Mark1}.
Nevertheless, the simulations are able to show some
important aspects, which can help to interpret the experimental results, as
will be shown below.

Experimentally, the maximum of the structure factor, which is the Fourier
transform of the monomer pair correlation function of all monomers in the
sample, shows a concentration dependent position 
$q_{max} \sim \rho _c ^{\nu_{sf}}$ with $\nu_{sf} = {\frac{1}{2}}$
at high densities and $\nu_{sf} = {\frac{1}{3}}$ at low densities, where
$\rho_c$ 
denotes the charge concentration.
Because the statistical fluctuations are too high, a reasonable distinction 
between the relevant exponents (0.5 to 0.3) can not be given in our case.
At low densities the scattering should reflect nothing but the
distances between the uniformly distributed chains. These distances vary with
$\rho ^{-1/3}$. At
higher densities the scattering results from the correlation hole that is
supposed to be proportional to the screening length which then is proportional
to the square root of the charge concentration \cite{Mark1,willi1,willi2}. 

For hydrophobic polyelectrolytes deviations of this behavior are found in
small angle neutron and x-ray scattering experiments \cite{willi2,spiteri}. An
effective exponent of $\nu_{sf}=0.4$ is observed, while the hydrophilic chains
show the usual exponent $\nu_{sf}= 0.5$. 
The investigated hydrophilic polymer is strongly charged and seems to
follows the Manning criterion over the examined small density range. So the
charge is ``fixed'', meaning the number of free charges in the
system remains constant. 
In the following we argue that the number of free charges in
the system is the relevant parameter to explain the
experimental results. 
In Figure \ref{ccplot} the number of condensed counterions
was plotted, which showed a strong density dependence. Because every
condensed counterion neutralizes effectively one charged monomer, this will
lead naturally to a reduced charged density at higher mass densities. The 
calculation of the screening length based on this ``effective'' charge density
reproduces the smaller exponents quite well. 
Only at the highest densities, where the number
of condensed counterions is almost constant, the exponent is given by the
volume change and equals $\nu_{sf}=0.5$. 
In Table \ref{tab:1} and \ref{tab:2} we estimate the effective hard sphere
diameter of the chains by calculating the DH radius. We assume that screening
is only facilitated by the mobile, non-condensed counterions with a density
$\rho_{fc}$.  This defines
a screening length $\kappa^{-1} = (4 \pi \lambda_B \rho_{fc})^{-1/2}$. If we
assume that the solution is in a liquid of hard spheres of radius
$\kappa^{-1}$ and if we further assume that the effective particles are
certainly big enough to interact, than we can argue that the effective
exponent $y$, obtained via $\kappa^{-1} \propto \rho ^y$ as given in Tables
\ref{tab:1} and \ref{tab:2} should reflect the 
experimental data for the scattering. There is however a word of care
needed. The value of $\nu_{sf} = 0.5$ as given from scaling assumes a blob
picture with strongly interacting chains. Thus this cannot expected to hold so
easily here. On the other hand for hydrophilic, shorter chains the crossover
from $\nu_{sf} = 1/3$ to $\nu_{sf} = 1/2$ was observed \cite{Mark1}.

Unfortunately quite often one considers in the experimental literature only 
the density of the chemical groups that can dissociate a counterion. 
We term this quantity the
``chemical'' charge density in this section, because it is the maximal charge
density possible for chemical reasons. The actual charge density will
almost always be smaller due to Manning condensation and additionally
condensation due to collapse. 

In Ref. \cite{willi1}, for example, 
a charge density of $f=0.04$ was found for a NaPSS molecule by osmotic
pressure measurements, with a chemical charge density given by
$f_{chem}= 0.38$. Although all analyzed systems are well above the critical
Manning value of one, and should therefore show a fixed
charge density, a strong variation of the measured charge densities was
observed. This is a clear
hint that strong condensation effects due to more compact conformations play an
important role for hydrophobic polyelectrolytes, in accordance to the
simulation results of section \ref{konf}. This has also consequences for the
osmotic pressure \cite{micka:phd}.

\begin{figure}[htb]
  \begin{center}
    \leavevmode
    \epsfig{file=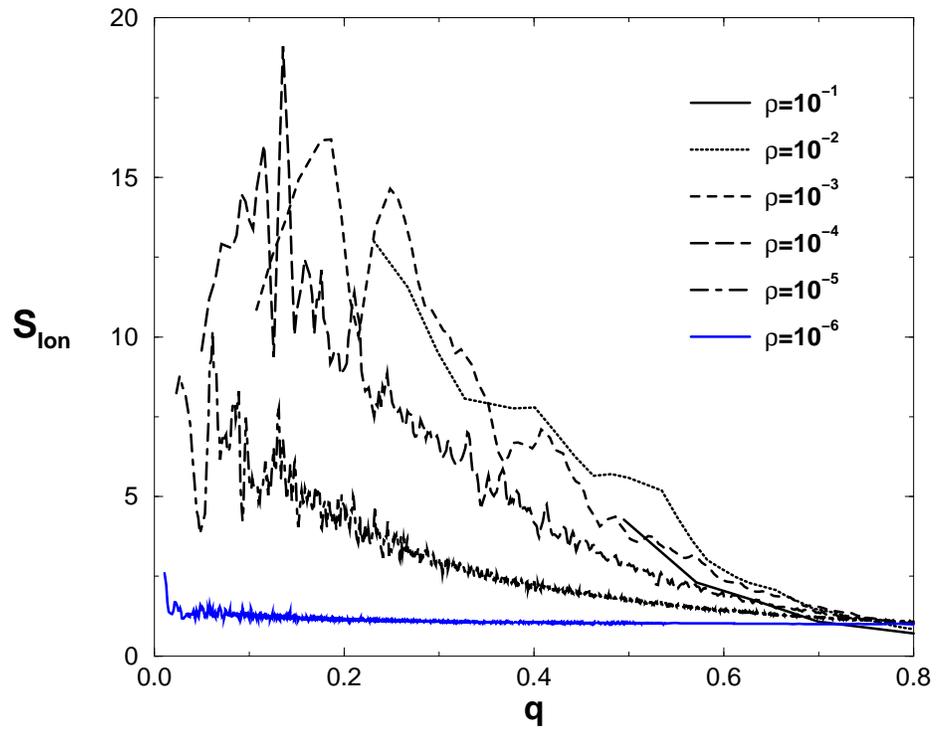, width=\textwidth}
    \caption{The structure factor $S_{ion}(q)$ vs. $q$ of the counter ion
    distribution for the indicated
    densities $\rho$. For lower densities the structure factor becomes similar
    to the one of a homogeneous background.}   
    \label{sfion1}
  \end{center}
\end{figure}

\begin{figure}[htb]
  \begin{center}
    \leavevmode
    \epsfig{file=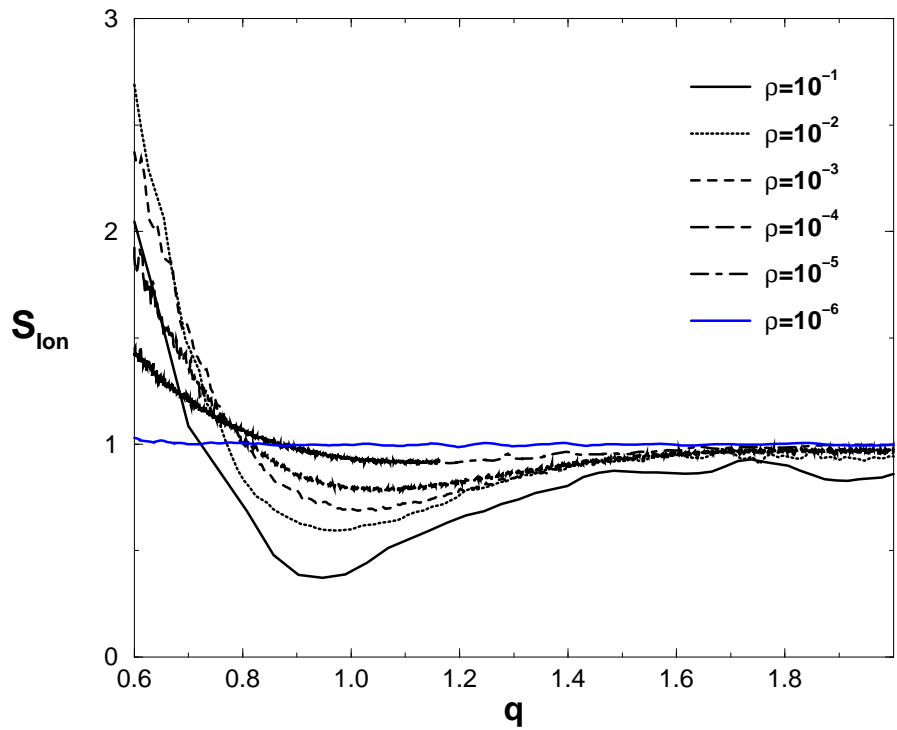, width=\textwidth}
    \caption{Blow up of Fig.~\ref{sfion1} for large $q$. One can detect a
    minimum of the structure factor for $q\approx1.0$, which is reminiscent of
    a correlation hole of the counterion cloud.}
    \label{sfion2}
  \end{center}
\end{figure}

The determination of the fraction of bound (immobile) counterions is very
difficult to accomplish experimentally. We imagine that particularly 
with x-ray and
synchrotron scattering for some well prepared systems one could
achieve a sufficient contrast between the counterions and the chains. As
the majority of the chain atoms are carbon, hydrogen and oxygen the use of
heavy counterions like cesium should give a sufficient contrast in
scattering intensities. Therefore the structure factor of the counterion
distribution should be measurable. This quantity is much easier calculated in
the 
simulation, because the counterions are more mobile and outnumber the number
of polymers by far.
So their distribution in the system can be dissolved in much
more detail. The simulations show very interesting results, as for example
in Figure \ref{sfion1} where the density dependency of the
small $q$ part of the counterion structure factor is shown. For high and medium
densities the distribution is coupled to that of the polymers and shows a
pronounced signal of the underlying structure. For the highest density, the
structure factor shows no clear 
peak. This is the reflection of the relatively uniform distribution of the
counterions in the microgel. In the medium density range a maximum shows up
which increases and moves to smaller $q$-values with decreasing density. In the
case of \ro{5} the intensity decreases, because more
and more counterions leave the strongly stretched chains so that a significant
background contributes to the scattering. For the lowest density only this
background of uniformly distributed counterions remains, leading to a
completely 
flat and unstructured scattering function. For higher $q$ a clear minimum is
detected, which is getting deeper with decreasing density, as can be seen in
Figure~\ref{sfion2}. Of course, the lowest density still gives the flat
signal from the uniform background. This minimum can be interpreted as a
correlation hole between the counterion clouds. At higher densities, many of
the counterions are condensed and therefore near the chains. The number of free
counterions is relatively small compared to the free volume between the chains
which spans almost the whole volume of the solution, as can be seen in Figure
\ref{poly2}. This concentration difference is reflected by the structure
factor. At even higher $q$-values a second maximum can be detected for the
higher densities in the simulation data. It is located around the $q$-value
related to the radius of gyration of the polyelectrolytes reflecting the
correlation of the chains within the counterion cloud around one monomer. 
As the amplitude is very small this effect is difficult to
analyze quantitatively, however we consider it not to be relevant for
experimental studies at present, and do not analyze it in further detail here. 
%
\section{Conclusions}
%
We present a set of MD simulations studying the influence of a poor solvent on
the structural properties of strongly charged, flexible polyelectrolyte chains
in salt free solution.

We showed that the underlying neutral system is
characterized by a $\theta$ point which we determined to be
$\epsilon_{LJ,\theta} = 0.34 \pm 0.02 $. In the beginning three different
values of the strength of the hydrophobicity ($\epsilon_{LJ} = 0.25$, 0.5, and
1.0) were examined at a polymer density of 
$\rho=10^{-4}\sigma^{-3}$. The value of $\epsilon_{LJ} = 0.25$ lead to
hydrophilic behavior as expected, because this value is even below the neutral
$\theta$ point. However, the other two values of $\epsilon_{LJ}$ showed as
well only a weak influence of the attractive interaction and the differences
to the purely hydrophilic chains was rather small. By varying the Bjerrum
length $\lambda_{B}$ however, we could demonstrate that under the right
conditions  weak hydrophobicities can nevertheless lead to strong effects. 
For $\epsilon_{LJ} = 0.5$ the chains showed a stretching followed by a strong
contraction with increasing Bjerrum length. This effect is well known from
simulations of hydrophilic polyelectrolytes, but in our hydrophobic case it is
more pronounced and shifted to smaller $\lambda_{B}$. 

An ``effective'' $\theta$ point was calculated for the charged system for
a density of $\rho = 10^{-4}\sigma^{-3}$. The condition that its end-to-end
distance equals the one for a ideal, neutral chain of the same
length resulted in $\epsilon_{LJ,\theta_{eff}}^{(c)} = 1.51 \pm 0.03$. 
This value was taken as a starting point for detailed study of the chain
conformations under a variation of the polymer density. At high densities we
found a high screening of the electrostatic interactions, favoring the short
range hydrophobic attraction, whereas at low densities the Coulomb forces 
clearly dominated.
A multitude of chain conformations
ranging from micro-gels via isolated, strongly collapsed chains to extremely
stretched necklace chains was found. The collapsed objects showed locally Porod
scattering ($S(q) \propto q^{-4}$), while the necklace like objects showed a logarithmic
slope of their scaling function of $m=-1$ ($\nu = 1$),
so that both extremes are realized in one system
that differs only by density. A micro-phase separation was not observed. 
We estimated from our simulation that an agglomeration of two chains is
unfavorable for the density \ro{2}, whereas more detailed estimates are
given in \cite{letter}.
To sum up, the polymer density seems to be a very useful parameter to
vary to study polyelectrolytes under poor solvent conditions. 
The preparation of such systems should be extremely fruitful also for 
experiments, because the density $\rho$ can be very easily controlled. The
comparison of the MD data of the 
structure factor of the monomer pair correlation with experimental data 
indicates that the knowledge
of the real number of free charges in the system is of central importance in
understanding the behavior of polyelectrolyte solutions, especially for the 
poor solvent case.  
Measurements of the counterion structure factor would be very helpful
in comparison to the simulation results presented here.

We could relate the structural properties of
polyelectrolytes under good and poor solvent conditions
within one consistent approach. This gave a
clear and conclusive picture of the influence of solvent quality on
the structure of flexible polyelectrolyte chains.
%
\section*{Acknowledgments}
%
We would like to acknowledge interesting and stimulating discussions with
M. Deserno, B. D\"unweg, T. Liverpool, and F. M\"uller-Plathe. 
A large grant of computer time at the HLRZ J\"ulich under grant hkf06 
is gratefully acknowledged.
%

\newpage
\begin{center}
{\bf \Large List of Abbrevations and Symbols}
\end{center}
\begin{list}{*}{
\setlength{\parsep}{0.0cm}
\setlength{\labelwidth}{2.1cm}
\setlength{\labelsep}{0.3cm}
\setlength{\leftmargin}{2.2cm}   }
\item[MD]  Molecular Dynamics     
\item[LJ]  Lennard-Jones    
\item[FENE]  finite extension nonlinear elastic
\item[PME]Particle-Mesh-Ewald
\item[PSS] sulfonated polystyrene
\item[$U_{LJ,hydrophil}$]  purely repulsive hydrophilic potential          
\item[$U_{LJ,hydrophob}$]  hydrophobic potential showing a short range
  attraction 
\item[$U_{FENE}$] Finite Extension Nonlinear Elastic bond potential between
  chain monomers
\item[$R_{end}$] end-to-end distance
\item[$R_G$] radius of gyration
\item[$r$] characteristic ratio $r = \frac{R^2_{end}}{R_G}$
\item[$N$] Numbers of monomers of a chain
\item[$L$] linear length of the simulation cell
\item[$b$] average bond length
\item[$\lambda_B$] Bjerrum length
\item[$\kappa$] Debye-H\"uckel parameter
\item[$e$] unit charge
\item[$q$] valence of a unit charge
\item[$f$] fraction of charged monomers
\item[$f_{chem}$] chemical charge density
\item[$Q$] critical charge determined by the Rayleigh criterion
\item[$E^{Coul}$] Coulomb energy
\item[$d$] distance between globules
\item[$k_B$] Boltzmann constant
\item[$T$] Temperature
\item[$\sigma$] Lennard-Jones length unit (monomer radius)
\item[$\tau$] Lennard-Jones time unit
\item[$\epsilon_0, \epsilon_S$] permittivity of the vacuum and the solvent,
  respectively 
\item[$\Gamma$] friction coefficient of the Langevin dynamics
\item[$\Delta t$] time step of the velocity Verlet MD algorithm
\item[$\xi_M$] Manning Parameter
\item[$\xi_M^*$] critical Manning parameter
\item[$\rho$] density: number of monomers per simulation volume
\item[$T^{(n)}_\theta$] $\theta$ - temperature of the neutral system
\item[$T^{(c)}_\theta$] $\theta$ - temperature of the charged system
\item[$\epsilon^{(n)}_{LJ,\theta}$] LJ energy parameter giving the $\theta$
  point of the neutral system
\item[$\epsilon^{(c)}_{LJ,\theta_{eff}}$] LJ energy parameter giving the
  effective $\theta$ - point of the charged system
\item[$\tilde T$] reduced temperature to the critical $\theta$ - point
\item[$S_{sp}$] sphericall averaged structure factor
\item[$S_\parallel$] component of the spherically averaged structure factor
  along the first principal axis of inertia
\item[$q$] magnitude of the Fourier space vector
\item[$q_{max}$] position of the maximum of the structure factor
\item[$\Pi$] osmotic pressure
\item[$\nu$] scaling exponent of the chain extension
\item[$\nu_{sf}$] scaling exponent of the maximum of the structure factor
\item[$\rho_c$] charge concentration
\item[$\rho_{fc}$] density of mobile, noncondensed counterions
\end{list}

\newpage   
\begin{table}[h]
\begin{center}
\renewcommand{\tabcolsep}{0.5cm}
\begin{tabular}[tb]{|c|r|r|r|r|r|}
\hline
\multicolumn{6}{c}{\rule[-3mm]{0mm}{8mm}\rule[1mm]{0mm}{2mm}Variation of
the effective screening for hydrophobic polyelectrolytes}\\ 
\hline
\hline
\multicolumn{1}{c|}{\rule[-3mm]{0mm}{8mm}$\rho [\sigma^{-3}]$} &
$\text{N}_{C}$ & $\text{N}_{ch}$ & $\rho_{ch} [\sigma^{-3}]$ &
$\kappa^{-1} [\sigma]$ & \multicolumn{1}{r}{$y$} \\ 
\hline
\hline
$2 \cdot 10^{-1}$ & 25,5& 208&$1.95\cdot10^{-2}$ & 1.17 &
\\ \cline{1-5}
$2 \cdot 10^{-2}$ & 26 & 192 &$ 1.85\cdot10^{-3}$ & 3.79 &
\raisebox{1.5ex}[-1.5ex]{0.51} \\ \cline{1-5}
$2 \cdot 10^{-3}$ & 24.5& 240 & $2.3\cdot10^{-4}$ & 10.67 &
\raisebox{1.5ex}[-1.5ex]{0.45}\\ \cline{1-5}
$2 \cdot 10^{-4}$ & 21 & 352 &$ 3.44\cdot10^{-5}$ & 27.75 &
\raisebox{1.5ex}[-1.5ex]{0.41} \\ \cline{1-5}
$2 \cdot 10^{-5}$ & 13& 608 & $5.93\cdot10^{-6}$ & 66.87 &
\raisebox{1.5ex}[-1.5ex]{0.38} \\ \cline{1-5}
$2 \cdot 10^{-6}$ & 3& 928&$ 9.06\cdot10^{-7}$& 171 & 
\raisebox{1.5ex}[-1.5ex]{0.41}\\ 
\hline
\hline
\end{tabular}
\end{center}
\caption{Variation of the effective screening with density for hydrophobic
  polyelectrolytes. $N_C$ denotes the number of condensed counterions, $N_{ch}$
  is the number of free charges, and  $\rho_{ch}$ is the density of free
  charges. The exponent $y$ is obtained via $\kappa^{-1} \propto \rho ^y$. One
  recognizes a deviation of $y=1/2$ due to counterion condensation}
\label{tab:1}
\end{table}
\begin{table}[h]
\begin{center}
\renewcommand{\tabcolsep}{0.5cm}
\begin{tabular}[tb]{|c|r|r|r|r|r|}
\hline
\multicolumn{6}{c}{\rule[-3mm]{0mm}{8mm}\rule[1mm]{0mm}{2mm}Variation of
the effective screening for hydrophilic polyelectrolytes}\\ 
\hline
\hline
\multicolumn{1}{c|}{\rule[-3mm]{0mm}{8mm}$\rho [\sigma^{-3}]$} &
$\text{N}_{C}$ & $\text{N}_{ch}$ & $\rho_{ch} [\sigma^{-3}]$ &
$\kappa^{-1} [\sigma]$ & \multicolumn{1}{r}{$y$} \\ 
\hline
\hline
$2 \cdot 10^{-1}$ & 24& 256&$2.4\cdot10^{-2}$ & 1.05 &
\\ \cline{1-5}
$2 \cdot 10^{-2}$ & 21 & 352 &$ 3.4\cdot10^{-3}$ & 2.79 &
\raisebox{1.5ex}[-1.5ex]{0.42} \\ \cline{1-5}
$2 \cdot 10^{-3}$ & 18& 448 & $4.3\cdot10^{-4}$ & 7.85 &
\raisebox{1.5ex}[-1.5ex]{0.45}\\ \cline{1-5}
$2 \cdot 10^{-4}$ & 16 & 512 &$ 5\cdot10^{-5}$ & 23.03 &
\raisebox{1.5ex}[-1.5ex]{0.47} \\ \cline{1-5}
$2 \cdot 10^{-5}$ & 11.5& 656 & $6.4\cdot10^{-6}$ & 64.38 &
\raisebox{1.5ex}[-1.5ex]{0.45} \\ \cline{1-5}
$2 \cdot 10^{-6}$ & 3& 928&$ 9.06\cdot10^{-7}$& 171 & 
\raisebox{1.5ex}[-1.5ex]{0.42}\\ 
\hline
\hline
\end{tabular}
\end{center}
\caption{Variation of the effective screening with density for hydrophilic
  polyelectrolytes. The notation is the same as in Table \ref{tab:1}.
  The variations are clearly much weaker than for the hydrophobic case.}
\label{tab:2}
\end{table}

\newpage
%
%
%
\newpage
\newpage
\newpage
\newpage
\newpage
\newpage

\newpage

\newpage
\newpage
\newpage
\newpage
\newpage
\newpage
\newpage
\newpage
\newpage
\newpage
\newpage
\newpage
\newpage
\newpage
\newpage
\end{document}